\documentclass[nolinenumber]{aa}  
\usepackage{graphicx}
\usepackage{txfonts}
\usepackage[
  breaklinks = true,
   colorlinks = true,
  urlcolor   = blue,
  citecolor  = blue,
  linkcolor  = red, ]{hyperref}
\usepackage{ulem}
\usepackage[rightcaption]{sidecap}
\usepackage{color}
\usepackage{siunitx}
\usepackage{placeins}
\newcommand{\dn}{DN\,s$^{-1}$\,px$^{-1}$}
\begin{document} 
\title{Radio, X-ray, and EUV signatures of internal and external reconnection of an erupting flux rope}

   \author{Jana Ka\v{s}parov\'{a}\inst{1}
            \and Jaroslav Dud\'{i}k\inst{1}
            \and Marian Karlick\'{y}\inst{1}
            \and Alena Zemanov\'{a}\inst{1}
            \and Paolo Massa\inst{2}
            \and Samuel Krucker\inst{2, 3}
            \and Fr\'ed\'eric Schuller\inst{4}
            \and J\'{a}n Ryb\'{a}k\inst{5}
          }

    \institute{Astronomical Institute of the Czech Academy of Sciences, Fričova 298, 25165 Ondřejov, Czech Republic
    \and Institute for Data Science, University of Applied Sciences and Arts Northwestern Switzerland, Bahnhofstrasse 6, Windisch, 5120, Switzerland
    \and Space Sciences Laboratory, University of California Berkeley, 7 Gauss Way, Berkeley, CA 94720, USA
     \and Leibniz-Institut für Astrophysik Potsdam (AIP), An der Sternwarte 16, 14482, Potsdam, Germany
    \and Astronomical Institute, Slovak Academy of Sciences, Tatransk\'{a} Lomnica, Slovakia
    }

  \abstract
   {}
   {We aim to interpret and relate features observed in GHz radio, extreme UV (EUV), and X-ray emissions during a filament eruption at the start of a solar flare on 2 April 2022.}
   {We analyse imaging (EUV, X-ray) and spectral (radio, X-ray) data obtained by ground based and space instruments on board space missions both on Earth 
   (Fermi, Hinode, Solar Dynamics Observatory) and solar orbit (Solar Orbiter, STEREO-A), which provide a multi-directional view on the same event.}
  {The combination of EUV and X-ray images and X-ray spectra allowed us to identify hot loops in the vicinity of the filament before its eruption. We interpreted their interaction with the rising filament as a signature of an arcade-to-rope reconnection geometry. The subsequent EUV brightening within the filament revealed helical structure of the erupting rope. We explained co-temporal 
  radio slowly positively drifting bursts  
  as a result of beam acceleration within the magnetic rope and propagation along the helical structure. Corresponding X-ray spectra were consistent with a thermal origin. 
  The filament rising was accompanied by co-temporal normal and reverse drift type III radio bursts. We interpreted them as a signature of a reconnection event 
  and estimated electron density at the reconnection site.
  Further untwisting of the helical structure led to formation of a 
  quasi-circular EUV structure seen from Earth and STEREO-A. 
  Its occurrence was co-temporal with a unique tangle of radio U- and inverse U-bursts. 
  We proposed that several accelerated beams propagate within that complex structure and generate the burst tangle. During the start of the flare hard X-ray emission was concentrated near the filament leg only suggesting predominant propagation of the beams towards its rooting.}   
   {
   We collected multi-wavelength observations indicating interaction of the erupting magnetic flux rope with the overlying arcade and internal magnetic reconnection inside the rising flux rope.
   }
   

   \maketitle
   \nolinenumbers
\section{Introduction}

Solar flares are the most powerful events in the the solar system, releasing energy of up to 10$^{32}$ ergs during several minutes or tens of minutes. The strongest flares are associated with coronal mass ejections and acceleration of particles into interplanetary space. From physical point of view solar flares are catastrophic phenomena in the solar atmosphere, in which the energy accumulated in the magnetic field and electric currents is rapidly transformed into plasma heating, plasma flows, accelerated particles and emission in a broad range of electromagnetic waves: from radio, through optical, UV, X-rays to gamma-rays \citep[see, e.g., the reviews by][]{2002A&ARv..10..313P,2002SSRv..101....1A,2008A&ARv..16..155K,2009AdSpR..43..739S,2011SSRv..159...19F,2016SSRv..200...75N}.  This conversion of the magnetic energy into kinetic energy and radiation is facilitated by the process of magnetic reconnection 
\citep[see, e.g.,][]{Priest00,Zweibel09,Janvier17,Pontin22,Dudik25},
during which the magnetic field changes its connectivity into a lower energy state. In three dimensions, a necessary condition for magnetic reconnection is the existence of electric field (and thus current) parallel to the magnetic field \citep[see][]{Hesse88,Schindler88,Janvier14}.

The most powerful flares are the eruptive flares.
These flares, often described by the so-called standard flare model, now generalised into 3-dimensions
\citep{Aulanier12,Janvier14}, employ a central element -- twisted, unstable magnetic flux rope, whose instability drives the CME eruption \citep[see, e.g.,][and references therein]{Amari2014,Fan2015,Zuccarello15}. Solar magnetic flux ropes are often observable as filaments and prominences when filled with relatively cool chromospheric plasma residing in its bottom parts in the magnetic dips \citep[see, e.g.,][]{Aulanier98,Aulanier99,Aulanier02,Dudik08}. Although some filaments or parts of filaments may correspond to a sheared arcade rather than a flux rope \citep[see e.g.][]{Guo10}, the magnetic configuration of a filament at its eruption typically resembles that of a magnetic flux rope \citep[see the review of][]{Patsourakos20}. During a flux rope eruption, the most energetic magnetic reconnection processes usually occur in the narrow current sheet formed below the rising flux rope, although the magnetic flux rope itself can also often reconnect with the surrounding corona
\citep[ar--rf reconnection geometry, see][]{Aulanier19,Dudik19,Gou23}.
This reconnection geometry leads to drifting of the footpoints of the erupting flux rope (called the "Aulanier effect"\footnote{\url{https://heliowiki.smce.nasa.gov/wiki/index.php/The_Aulanier_Effect:_drifting_footpoints_of_CME_flux_ropes}}) as the flux rope changes its connectivity.

Eruptive and flaring processes following magnetic reconnection are well-observed in the EUV and soft X-ray (SXR) wavelengths emitted by the flaring solar corona. The post-reconnection plasma is typically observed to be hot, with temperatures above 10\,MK, and up to several tens of MK 
\citep[the literature is extensive, but see, e.g.,][]{Feldman80,2011SSRv..159...19F,Warren11,Dudik14,Hudson21}.
Such hot plasma is not observed only in the form of the arcade of flare loops, bright in both EUV and SXR, but also in the form of hot ejecta moving upwards and associated with the erupting flux rope \citep[e.g.,][]{Zhang12,Su13,Liu13,Patsourakos13,Dudik19}, sometimes also visible at radio wavelengths \citep{Huang19,Chen20,Carley20}. Reconnection involving the erupting flux rope have recently been linked to significant heating of the filament plasma, which also strongly brightened as a result \citep{Joshi2025}.

Presence of accelerated particles produced by magnetic reconnection is routinely detected in the form of hard X-ray (HXR) emission. While SXR is believed to be produced by a hot plasma owing to the thermal bremsstrahlung  
\citep{1988psf..book.....T}, HXR emission is generated by electron beams via the non-thermal bremsstrahlung \citep{1971SoPh...18..489B}. The HXR sources are found typically at flare footpoints, although HXR coronal sources are common as well \citep[e.g.][]{2011SSRv..159...19F}. Recently, non-thermal HXR sources were found at the footpoints of erupting filaments \citep{Stiefel2023,Purkhart23}.

In radio, flares are accompanied by several types of bursts (type II, III, IV, V, J, U) and their fine structures \citep{1985srph.book.....M,2004psci.book.....A}. 
Catalogues of these bursts and fine structures in the decimetric range are shown in \cite{1994A&AS..104..145I,2001A&A...375..243J}. The most frequent radio bursts are type III bursts, occurring mostly during the flare impulsive phase. It is commonly accepted that they are generated by beams of accelerated electrons through the plasma emission mechanism  \citep{1980panp.book.....M}. They are observed with the normal (negative) or reverse (positive) frequency drift. Negative drift means that the electron beam propagates upwards in the solar atmosphere towards the lower plasma density; while positive drift signifies movement of accelerated electrons towards the higher plasma density, see also \citet{2020FrASS...7...56R}. Type U, J and V bursts can be considered as variants of type III bursts. While types U and J bursts are generated by beam electrons propagating along curved loops \citep{2023A&A...669A..28M},
type V bursts are explained by beam electrons propagating as in the type III burst case and then trapped at some location in the coronal loop  \citep{2013ApJ...779...83T}. Other bursts frequently observed during flares are type II and type IV bursts. Type II bursts are produced by flare shocks \citep{2021ApJ...923..255K,2022A&A...660A..71M} and type IV bursts (also called continua) are proposed to be generated by supra-thermal electrons trapped at the upper parts of flare loops \citep{2011fssr.book.....C,2023Ge&Ae..63..153F}. Besides these well known types of radio bursts there are the drifting pulsation structures (DPS) observed at the very beginning of the eruptive flares in connections with the plasmoid located usually at the upper part of the flux rope 
\citep{2000A&A...360..715K,2008SoPh..253..173B}. 

Except some papers \citep{Huang19,2020ApJS..250...31K} the energetic magnetic reconnection processes within the rising magnetic rope itself are not considered. Another phenomenon, that could be caused by the magnetic reconnection inside the rising magnetic rope, is the unique negatively drifting continuum observed in the 6 September 2017 flare \citep{2020ApJ...888...18K}.

But commonly it is believed that although the electric current inside the magnetic rope is huge, the electric current density (and the corresponding electron-ion drift velocity), which is essential for flare processes based on the anomalous resistivity, is very low owing to the large cross-section of the magnetic rope. Nevertheless, if the magnetic rope is structured, for example as a result of the rope formation before its ejection or owing to instabilities in its rising stage, then the current density and thus the electron-ion drift velocity at some locations in the magnetic rope could be enhanced. Thus, at these locations, reconnections within the flux rope, between its different parts, become possible. In such cases, these internal flux rope reconnections ought to be accompanied by the signatures of flaring activity, namely the presence of hot plasma emission in EUV and SXR, as well as accelerated particles producing emission in the HXR and radio parts of the spectrum.

In this paper, we present evidence of external reconnection in the arcade-to-rope geometry as well as internal reconnection of the erupting flux rope. 
Hot plasma above and within the flux rope was detected in EUV and SXR, HXR emission was concentrated near the leg of the rising filament, and multiple radio bursts, indicating particle acceleration and their motion along curved trajectories, were observed in GHz range. Some of these bursts were quite exceptional and we identified their EUV counterparts.

\begin{figure*}
    \centering    
    \includegraphics[width=6.36cm,viewport = 3 40 399 258,clip]{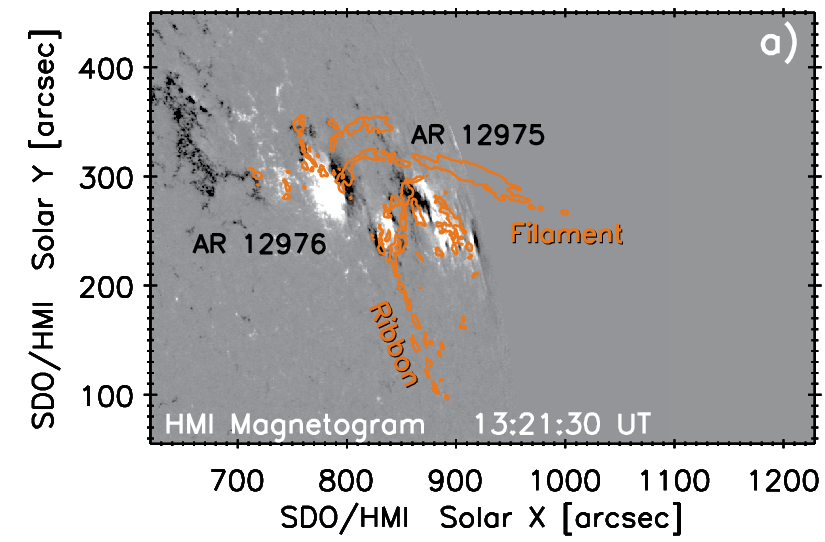}
	\includegraphics[width=10.64cm,viewport= 0  0 793 258,clip]{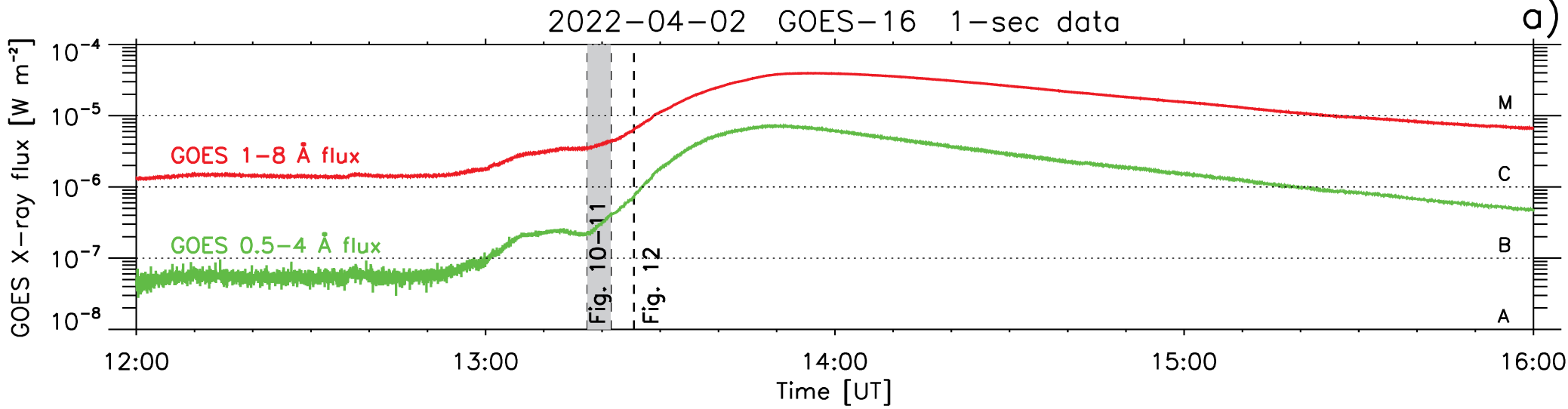}
    \includegraphics[width=6.36cm,viewport =  0 40 395 365,clip]{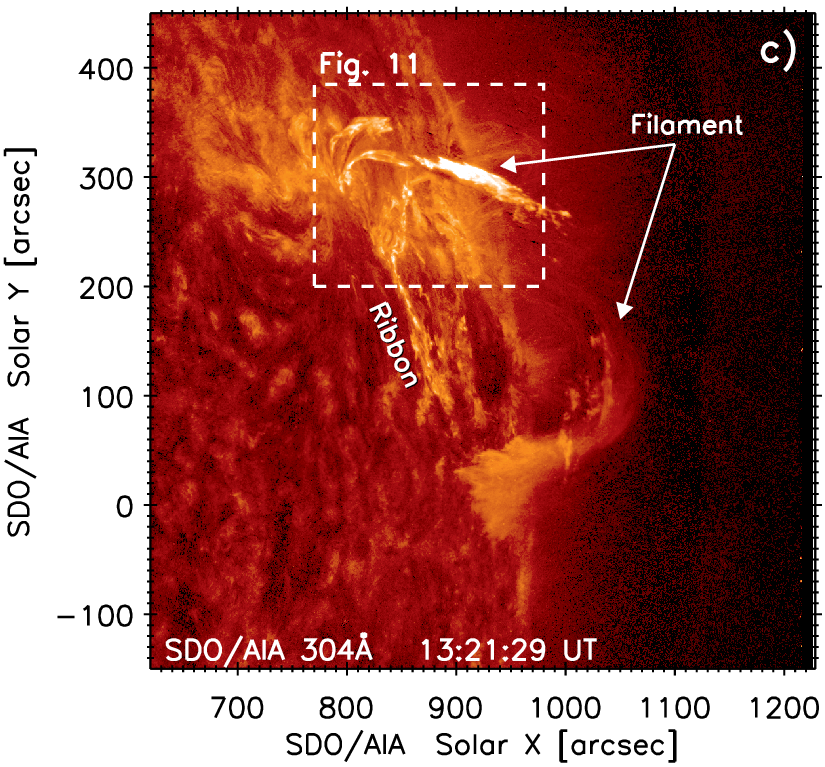}
    \includegraphics[width=5.32cm,viewport = 65 40 395 365,clip]{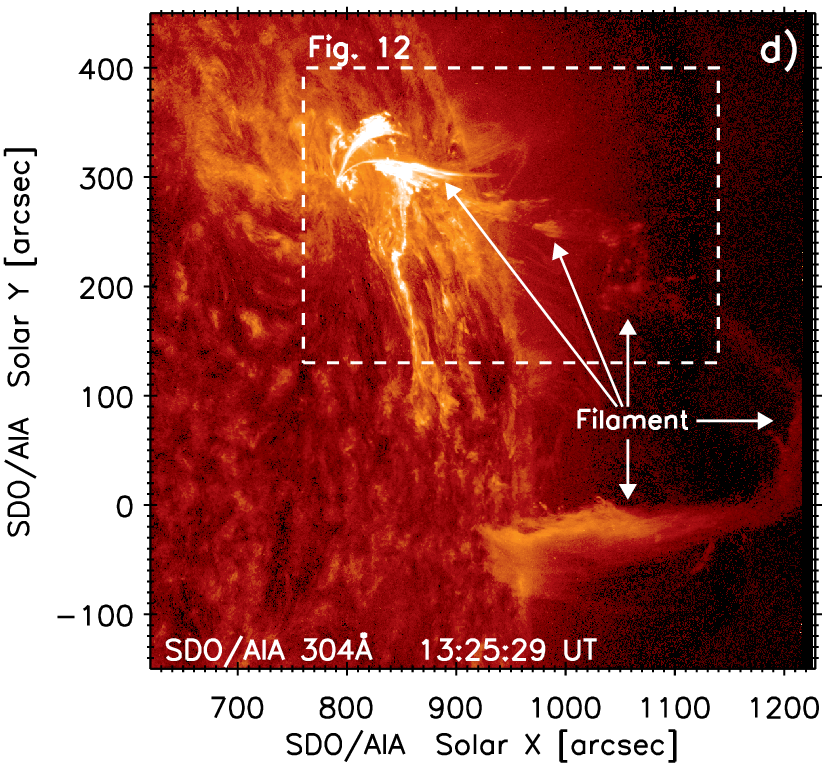}
    \includegraphics[width=5.32cm,viewport = 65 40 395 365,clip]{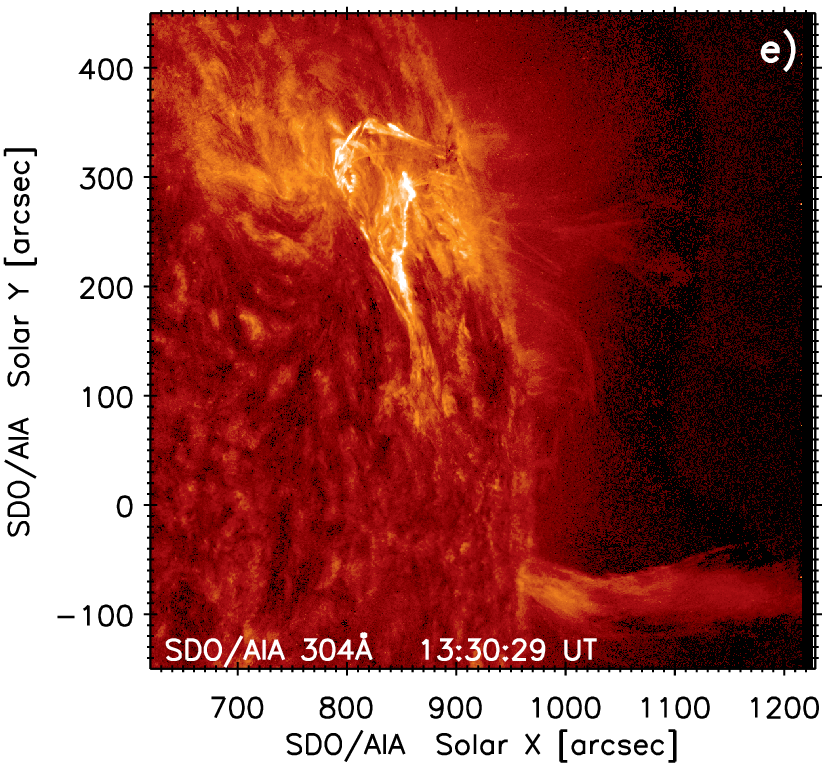}
    \includegraphics[width=6.36cm,viewport =  0  0 395 365,clip]{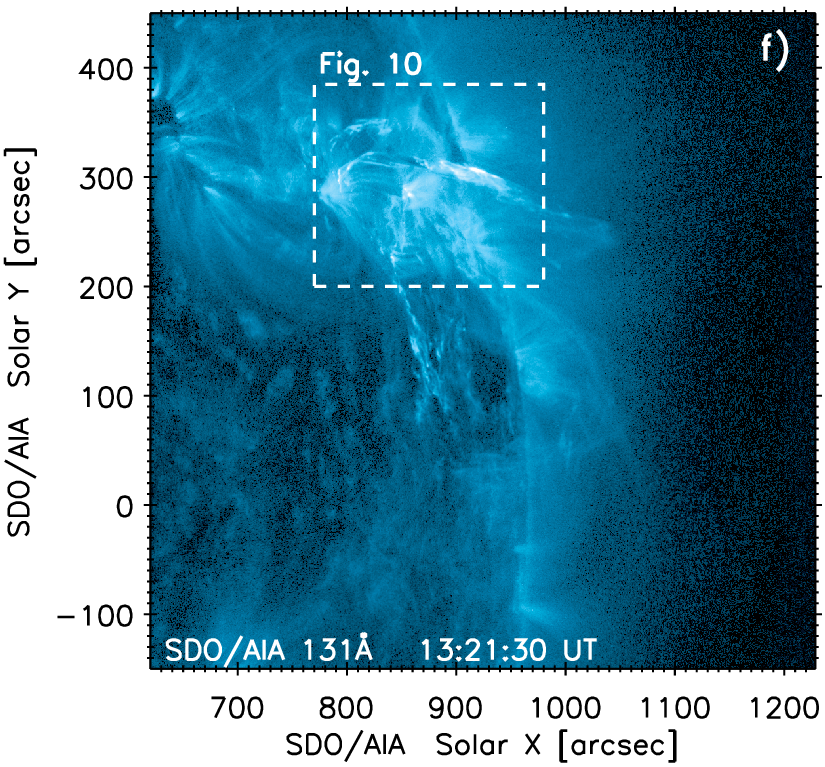}
    \includegraphics[width=5.32cm,viewport = 65  0 395 365,clip]{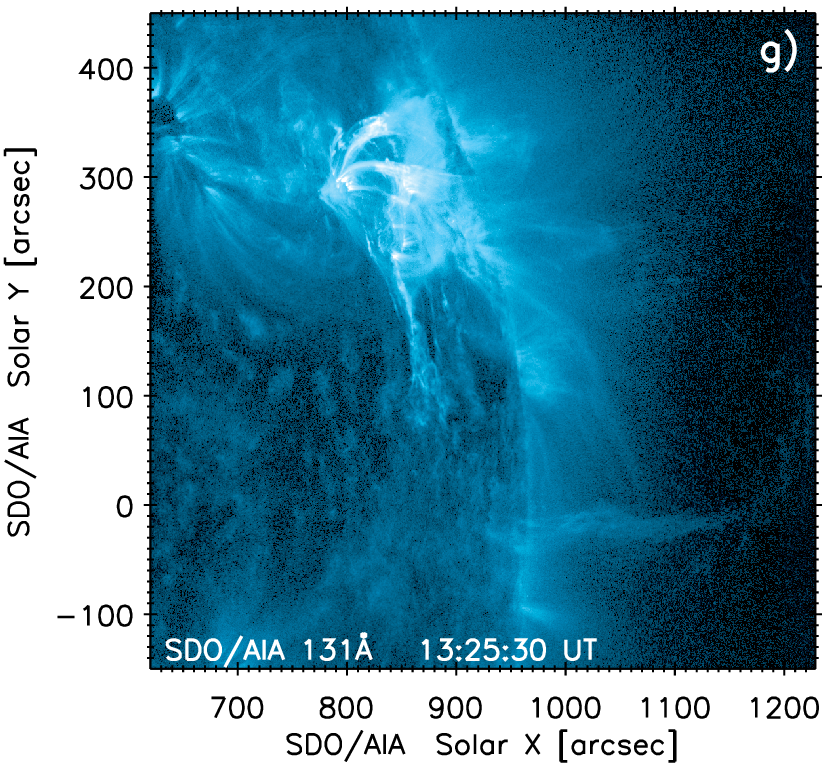}
    \includegraphics[width=5.32cm,viewport = 65  0 395 365,clip]{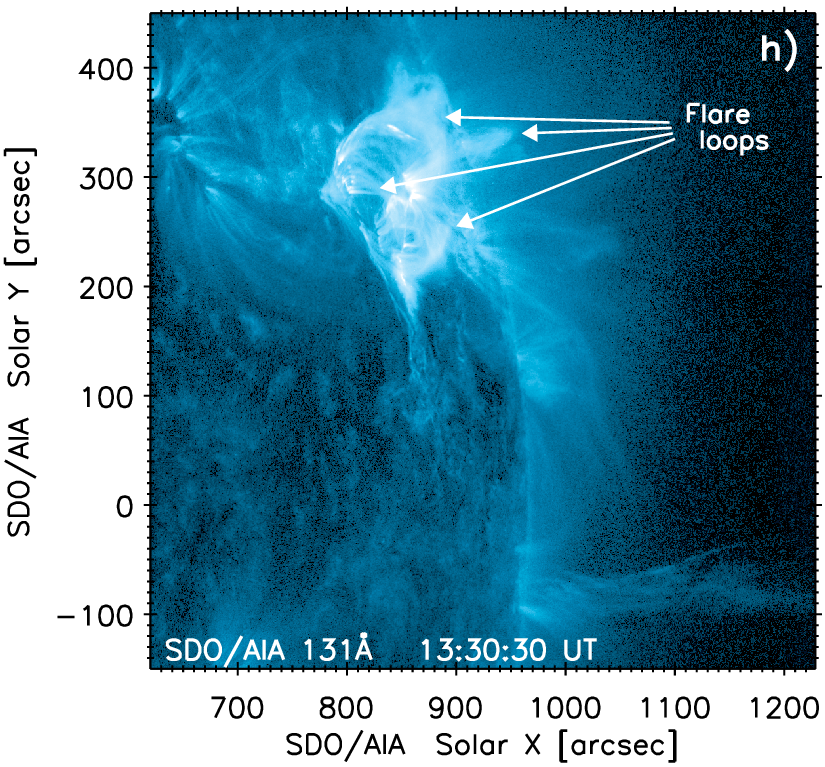}   
    \caption{Overview of filament eruption and accompanying flare. (a): SDO/HMI 
    magnetogram (b): GOES X-ray observations. (c)--(h): SDO/AIA images in 304\,\AA\, (c--e) and 131\,\AA~passbands (f--h). The filament and flare loops are indicated. The contour in panel (a) corresponds to 50~\dn\, in 304\,\AA~(panel c), showing the relative location of the erupting filament with respect to the photospheric polarities.   
    Dashed and dotted lines indicate times and FOV of Figs.~\ref{Fig:Hot_loop}, \ref{Fig:Hot_loop_304}, \ref{Fig:euv_1325UT}.
    An animation of the SDO/AIA data is available online, spanning 12:50–13:55~UT. 
    }
    \label{Fig:aia_overview}
\end{figure*}
\section{Eruption of 2 April 2022}
\label{Sect:2}

On 2 April 2022, an eruptive M3.9 flare, SOL2022-04-02T12:56:00L084C078\footnote{obtained from \url{https://www.lmsal.com/heksearch/}},  
occurred in a complex of active regions NOAA  12975--12976, located near the western limb of the Sun when observed from Earth (Figs.~\ref{Fig:aia_overview} and~\ref{Fig:HXR_evolution}). According to the GOES soft X-ray observations (Fig.~\ref{Fig:aia_overview}b), the flare started at 12:56 UT, reached its maximum about an hour later at 13:55 UT, and then followed a long decay phase. It is thus a long-duration event.

The event was well-observed using many instruments working across the electromagnetic spectrum. Global overview of the radio and hard X-ray lightcurves (observed by Fermi/GBM and Solar Orbiter/STIX) are shown in Fig.~\ref{Fig:HXR_evolution}, which serves as a guide to the many peaks in both spectral regions and their time of occurrence. EUV observations of the eruption and the accompanying flare are presented in Fig.~\ref{Fig:aia_overview}. It is seen that the event involves eruption of a relatively-long, transequatorial filament, whose northern footpoint was located in AR 12976
when viewed from the Earth (Fig.~\ref{Fig:aia_overview}a, c, d). The evolution of the photospheric magnetic field and its relation to the filament location is discussed in detail by \citet{Janvier23}.
The eruption was accompanied by a flare consisting of many systems of hot flare loops (Fig.~\ref{Fig:aia_overview}f--h).

The radio, EUV, and X-ray observations are further described in detail in Sects.~\ref{Sect:Radio_obs}, \ref{Sect:EUV_obs}, and \ref{Sect:X_obs}, respectively. The instruments detected the event at various distances from the Sun, which results in different onboard times, see Table~\ref{Table2} and Fig.~\ref{Fig:Spacecrafts}. The base time used throughout the text is the time of the event detection on Earth orbit in UT, otherwise the term onboard is specified.
\section{Observations}
\subsection{Radio spectra}
\label{Sect:Radio_obs}

The radio spectra were obtained using four different ground-based radiospectrographs:
Greenland-Callisto radiospectrograph working in the 10--100 MHz range with the resolutions of 0.25 s and 0.48 MHz \citep{2016JSARA..11...34M}, ORFEES radiospectrograph in the 150--1000 MHz range and the resolutions of 1.0 s and 0.98 MHz \citep{2021JSWSC..11...57H}, and two  Ond\v{r}ejov radiospectrographs \citep{2008SoPh..253...95J} in the ranges 800--2000 and 2000--5000 MHz with the time resolution of 0.01 s and frequency sampling of 4.7 MHz and 11.7 MHz, respectively.

\begin{table*}[!ht]
  \caption{Basic parameters of radio bursts identified in Figs.~\ref{Fig:radio_spdb} to \ref{Fig:radio_1324_1325}. 
  } 
\label{Table1}
\centering
\begin{tabular}{ccccccc}
  \hline\hline
Burst& Start & Duration & Range    & Type      & Frequency drift  & Figure         \\
 No. & (UT)  & (s)      & (GHz)    &           & (MHz s$^{-1}$) & No. \\
  \hline
\textbf{1}  & \textbf{13:21:11} & \textbf{3.4} & \textbf{1.5--1.8} & \textbf{SPDB}  &  \textbf{$\sim$+97}  & Fig.~\ref{Fig:radio_spdb} \\
\textbf{2}  & \textbf{13:21:30} & \textbf{4.2} & \textbf{1.5--1.8} & \textbf{SPDB}  &  \textbf{$\sim$+77} & Fig.~\ref{Fig:radio_spdb} \\
3  & 13:21:22 & 27      & 0.8--1.15& continuum & -  & Fig.~\ref{Fig:radio_spdb} \\
4  & 13:22:43 & 13      & 1.0--3.0 & continuum & -   & Fig.~\ref{fig5} \\
5  & 13:22:56 & 10      & 0.8--2.0 & continuum & - & Fig.~\ref{fig5}\\
6  & 13:23:13 & 17      & 0.8--1.4 & continuum & -  & Fig.~\ref{fig5}\\
7  & 13:23:00 & 3       & 2.7--4.2 & RS & $\sim+450$ & Fig.~\ref{fig5}\\
8  & 13:23:08 & 11      & 2.3--4.0 & hf-III, RS & $\sim-$1240, $\sim+$150, $\sim$+750 & Fig.~\ref{fig5}\\
9  & 13:23:30 & 33      & 0.8--1.3  & continuum  & -   & Fig.~\ref{Fig:radio_1323_1324}  \\
10 & 13:24:04 & 20      & 0.8--1.2  & continuum  & -  & Fig.~\ref{Fig:radio_1323_1324}   \\
11 & 13:23:49 & 4       & 1.3--1.8  & hf-III &  $\sim -1000$  & Fig.~\ref{Fig:radio_1323_1324} \\
12 & 13:24:06 & 4       & 1.3--1.8 & hf-III & $\sim -250$  & Fig.~\ref{Fig:radio_1323_1324}\\
13 & 13:23:51 & 4       & 2.7--4.5 & RS  & $\sim +800$ & Fig.~\ref{Fig:radio_1323_1324}\\
14  & 13:23:56 & 18      & 2.5--4.0 & hf-III, RS, complex & $-$360 to +870 & Fig.~\ref{Fig:radio_1323_1324}\\
15 & 13:24:18 & 6       & 2.7--3.7 & RS &  $\sim$+350 & Fig.~\ref{Fig:radio_1323_1324}\\
16 & 13:24:34 & 3       & 2.9--4.2 & U  & $-$2000 to +170   & Fig.~\ref{Fig:radio_1323_1324} \\
\textbf{17} & \textbf{13:25:05} & \textbf{4} & \textbf{1.3--2.0} & \textbf{RS, complex}  & \textbf{$+$100 to +500} & Fig.~\ref{Fig:radio_1324_1325}\\
\textbf{18} & \textbf{13:25:08} & \textbf{6} & \textbf{0.8--1.35} & \textbf{hf-III} & \textbf{$\sim$ $-$90} & Fig.~\ref{Fig:radio_1324_1325}\\
19 & 13:25:22 & 6       & 0.8--1.4 &  hf-III      & $\sim$-110& Fig.~\ref{Fig:radio_1324_1325}\\
\textbf{20} & \textbf{13:25:17} & \textbf{20} & \textbf{1.6--2.0} & \textbf{tangle of U bursts} & \textbf{$-$30 to +60} & Fig.~\ref{Fig:radio_1324_1325}\\
21 & 13:24:47 & 16      & 2.5--4.3 & group RS & $\sim$+1000 & Fig.~\ref{Fig:radio_1324_1325}\\
\textbf{22} & \textbf{13:25:18} & \textbf{20} & \textbf{2.0--2.8} & \textbf{tangle of inv-U bursts} & \textbf{$-$90 to +210} & Fig.~\ref{Fig:radio_1324_1325}\\
\hline
\end{tabular}
\tablefoot{
hf-III means the high frequency type III burst (negative frequency drift), RS is the type III burst with the positive (reverse) frequency drift and SPDB is the burst with very slow positive drift.
  U means U-burst and inv-U is the inverted U-burst. The most interesting  bursts (1, 2, 17, 18, 20, and 22), shown in bold, are analysed in detail.
}
\end{table*}
\begin{figure}[!ht]
    \centering
    \includegraphics[width=8.4cm,viewport=15 30 408 190,clip]{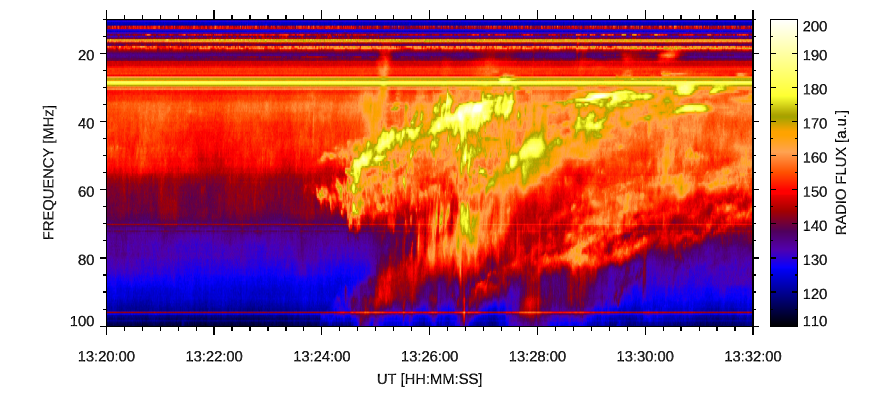}
    \includegraphics[width=8.4cm,viewport=15 30 408 190,clip]{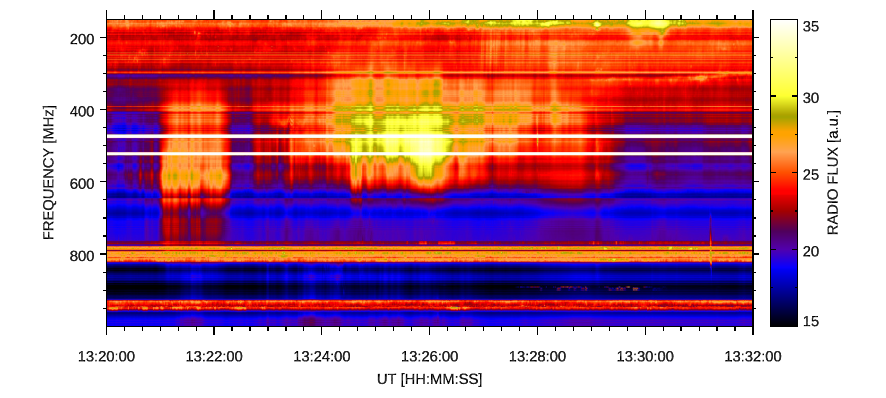}
    \includegraphics[width=8.4cm,viewport=15 30 408 190,clip]{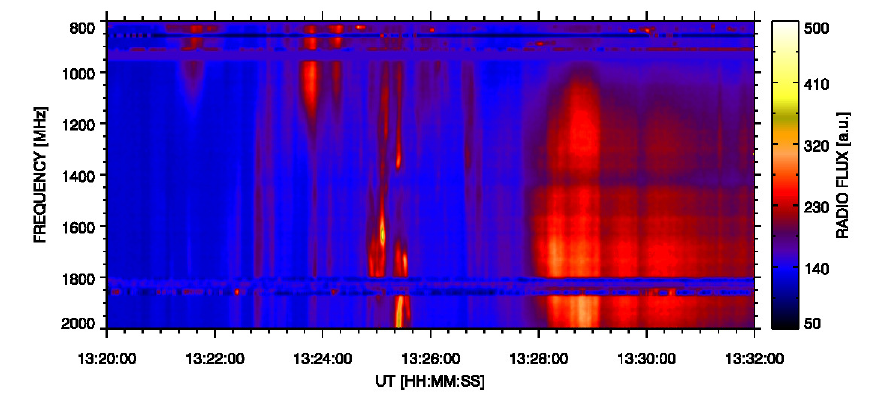}
    \includegraphics[width=8.4cm,viewport=15  0 408 190,clip]{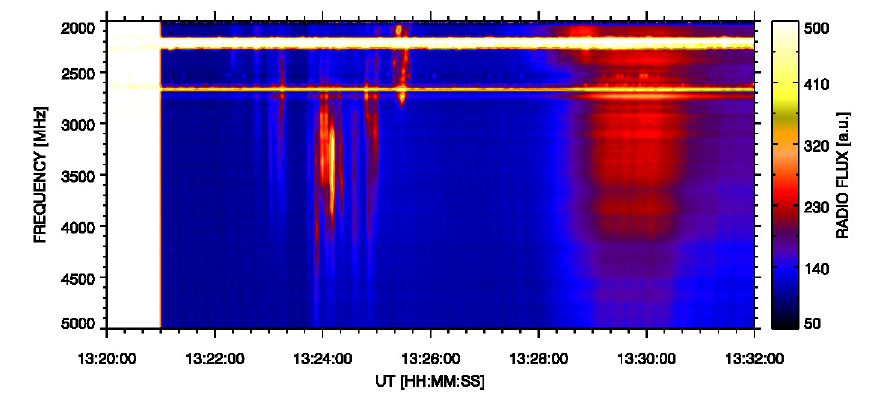}
    \caption{Overview of 20--5000 MHz radio spectrum in time interval 13:20-13:32 UT observed during 2 April 2022 flare. Top to bottom: 10--100 MHz Greenland-Callisto spectrum, 150--1000 MHz ORFEES spectrum, 800--2000 and 2000--5000 MHz Ond\v{r}ejov spectra. Horizontal emission bands are due to terrestrial artificial sources.}
    \label{Fig:radio_overview}
\end{figure}

An overview radio spectrum in the 20--5000 MHz range obtained by the above mentioned radiospectrographs is presented in Fig.~\ref{Fig:radio_overview}.
The radio flare started at about 13:21--13:22~UT in the 300--700~MHz range with the type III bursts. They were followed by pulsations recorded in the 300--600 MHz range at about 13:23--13:29 UT. In the 30--100~MHz range the type II burst with fundamental-harmonic branches was registered at 13:24--13:32~UT.

In the 800--5000 MHz range the radio emission started with a group of fast drifting bursts in the 13:21--13:28\,UT time interval, followed by a broadband continuum that was associated with the hard X-ray flux enhancement (Fig.~\ref{Fig:HXR_evolution}).  
 
From the point of view of the present study the most interesting bursts are in the 800--5000 MHz range. Therefore, in the following we show detailed spectra of these bursts in the 13:21:00--13:25:40\,UT time interval. Note that while these bursts in the overview spectrum look simple, their detailed spectra reveal their complexity.

\begin{figure}[!ht]
    \centering
    \includegraphics[width=8.4cm,viewport= 15 0 395 190,clip]{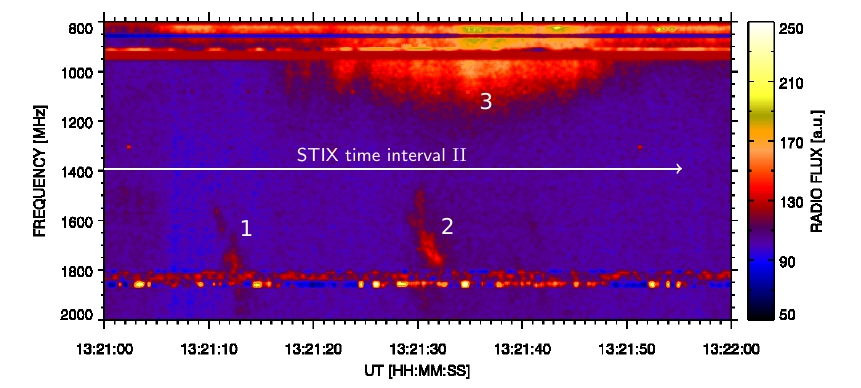}
    \caption{Detail of 800--2000 MHz radio spectrum in time interval 13:21:00--13:22:00 UT 
    at very beginning of flare. Numbers denote bursts (Table~\ref{Table1}), white horizontal bar denotes the STIX time interval II (Table~\ref{tab:stix_time_intervals}).}
    \label{Fig:radio_spdb}
\end{figure}

\begin{figure}[!ht]
    \centering
    \includegraphics[width=8.4cm,viewport= 15 30 408 190,clip] {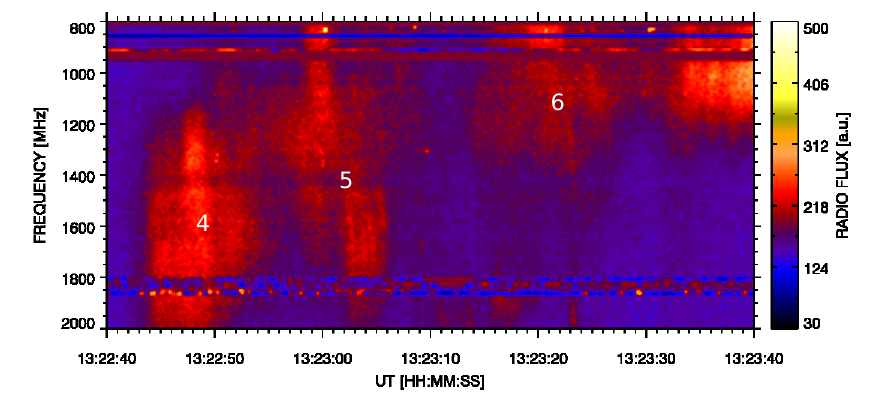}
    \includegraphics[width=8.4cm,viewport= 15  0 408 190,clip] {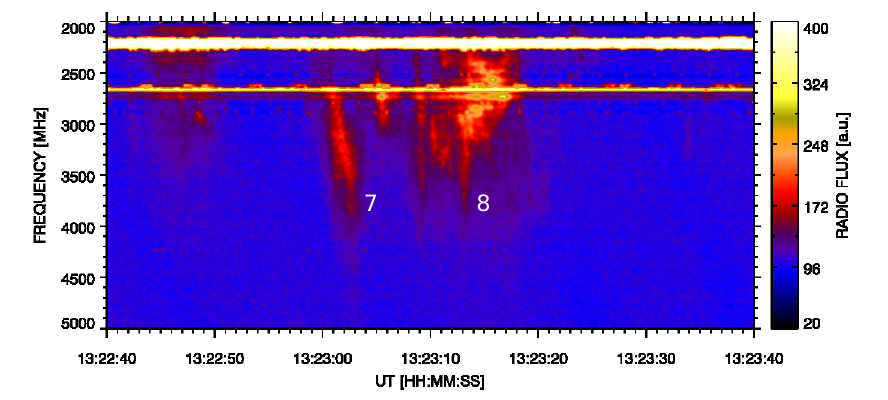}
    \caption{Details of 800--2000 and 2000--5000 MHz radio spectra in time interval 13:22:40--13:23:40 UT. Numbers denote bursts (Table~\ref{Table1}).}
    \label{fig5}
\end{figure}

\begin{figure}[!h]
    \centering
    \includegraphics[width=8.4cm,viewport=15 30 408 190,clip] {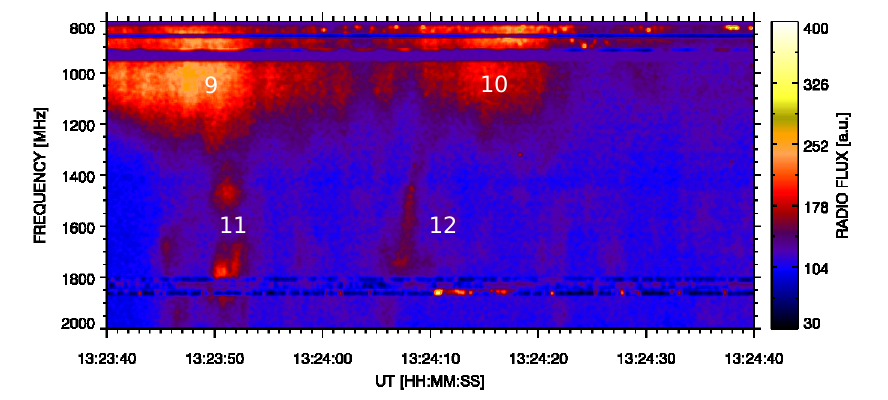}
    \includegraphics[width=8.4cm,viewport=15  0 395 190,clip] {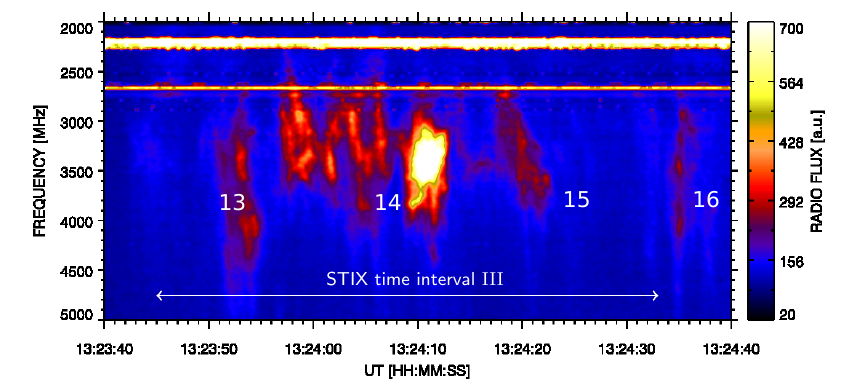}
    \caption{Details of 800--2000 and 2000--5000 MHz radio spectra in time interval 13:23:40--13:24:40 UT. Numbers denote bursts (Table~\ref{Table1}), white horizontal bar denotes the STIX time interval III (Table~\ref{tab:stix_time_intervals}).}
    \label{Fig:radio_1323_1324}
\end{figure}

\begin{figure}[!ht]
    \centering
    \includegraphics[width=8.4cm,viewport=15 30 408 190,clip]{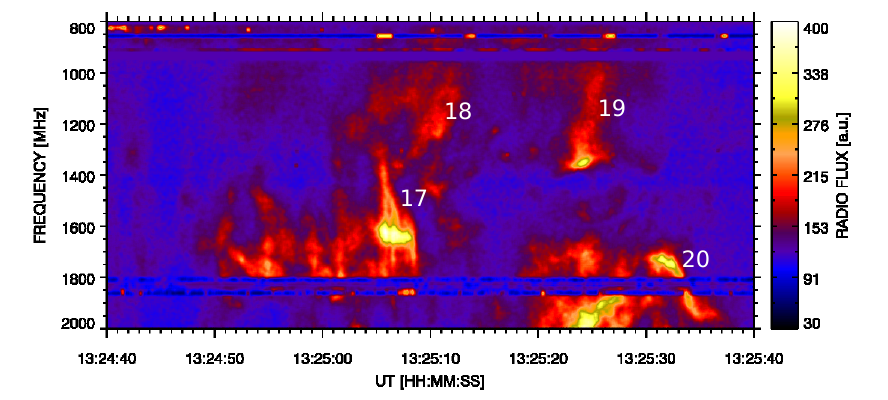}
    \includegraphics[width=8.4cm,viewport=15  0 395 190,clip]{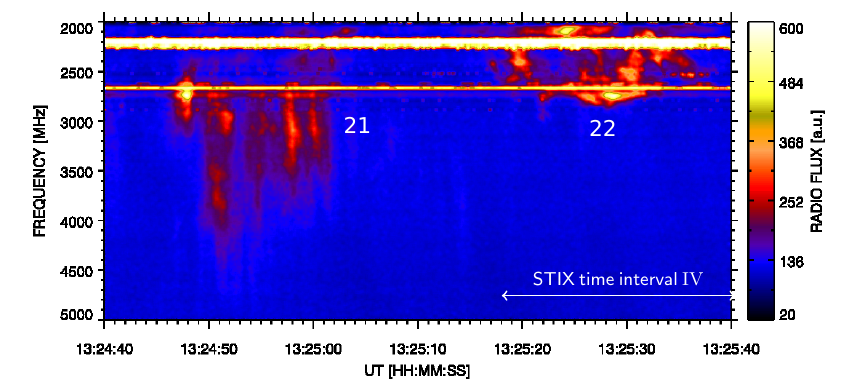}
    \caption{Details of 800--2000 and 2000--5000 MHz radio spectra in time interval 13:24:40-13:25:40 UT. Numbers denote bursts (Table~\ref{Table1}), white horizontal bar denotes the STIX time interval IV (Table~\ref{tab:stix_time_intervals}).}
    \label{Fig:radio_1324_1325}
\end{figure}

The 800--5000 MHz part of the radio spectrum observed at Ond\v{r}ejov represents the most interesting part of the radio spectrum, since it contains several rare or indeed unique radio bursts. Detailed radio spectra in the 800--5000 MHz range are shown in Figs.~\ref{Fig:radio_spdb} to \ref{Fig:radio_1324_1325}. The bursts in this frequency range are labelled using numbers 1--22. Their basic parameters: starting time, duration, frequencies, type, and frequency drift are summarised Table~\ref{Table1}. Most of these bursts are high frequency type III bursts (hf-III), the reverse drift bursts (RS) and continua. The both hf-III and RS are believed to be generated by electron beams, propagating upwards in the solar atmosphere (the case of hf-III) or downwards (the case of RS). Also type U and J bursts are assumed to be generated by electron beams, but propagating along the curved trajectory in the gravitationally stratified solar atmosphere \citep{2004psci.book.....A}. 

Besides these well-known types of bursts, we detected rarely observed slowly positively drifting bursts (SPDBs); see bursts 1 and 2 in Table~\ref{Table1}. This type of burst has been described and interpreted in \cite{2007SoPh..240..121F,Karlicky18}. In \cite{Karlicky18}, an SPDB was associated with a bright blob falling along a previously cold EUV structure, whereas in \cite{2020ApJ...905..111Z}, another SPDB was interpreted as being caused by a particle beam propagating along a helical magnetic structure within a nearly horizontally oriented magnetic rope.
The bursts 17 and 18 are also rarely observed together. They start at approximately the same frequency, but while burst 17 drifts to higher frequencies, burst 18 drifts to lower ones.
Finally, bursts 20 and 22 are entirely unique. No similar bursts have been recorded by the Ond\v{r}ejov radiospectrographs during their 30-year observational period. In the radio spectrum, they appear as a tangle of U- and inverted U-bursts.

\subsection{Imaging observations in EUV and soft X-rays}
\label{Sect:EUV_obs}
The filament eruption and the accompanying flare is well observed in EUV and X-ray parts of the spectrum by a multitude of instruments.
\subsubsection{SDO/AIA}
\label{Sect:AIA}
EUV imaging observations with high cadence are provided by the Atmospheric Imaging Assembly onboard Solar Dynamics Observatory \citep[SDO/AIA,][]{Lemen12} from the vantage point of Earth. The AIA instrument images the solar atmosphere in 10 EUV and UV filters at a cadence of up to 12\,s and spatial resolution of 1$\farcs$5, with a pixel size of 0$\farcs$6. The AIA data were processes using standard \texttt{aia\_prep} routine in SSWIDL. Residual spatial misalignments between some channels of about 1 pixel were corrected manually.

The AIA 304\,\AA~channel images the transition-region plasma in the \ion{He}{ii} 303.8\,\AA~line, and is useful for studying the dynamics of erupting filaments. In addition to the 304\,\AA~channel, the filament is seen as a dark absorbing structure in the coronal channels of AIA, where its optical depth and morphology are comparable to that of H$\alpha$ \citep{Anzer05}. The coronal channels we examined are primarily those of 171\,\AA, 193\,\AA, and 211\,\AA, which contain strong contributions from \ion{Fe}{ix}, \ion{Fe}{xii}, and \ion{Fe}{xiv} emission lines, respectively \citep{ODwyer10,DelZanna13}. We note that the AIA 171\,\AA~channel can contain filament-related \ion{Fe}{ix} emission, which is formed in the prominence-to-corona transition region at temperatures of above 400 kK \citep{Parenti12}.

The hot flare emission is observed in the 131\,\AA~channel, which in flares is dominated by the \ion{Fe}{xxi} 128.75\,\AA~line, formed at temperatures above log($T$\,[K])\,=\,7.0 \citep[][]{ODwyer10}. In quiescent coronal conditions, this channel is dominated by \ion{Fe}{viii} emission instead. We note that the morphology of the 131\,\AA~and 171\,\AA~passbands are in coronal conditions quite similar, since the \ion{Fe}{viii} and \ion{Fe}{ix} ions are formed close in temperature  \citep{ODwyer10,DelZanna13}. Therefore, presence of strong emission in 131\,\AA~but not in 171\,\AA~signifies that such plasma is hot, i.e. emitted by \ion{Fe}{xxi}, and originating at flare temperatures.

To visualise the hot plasma, we constructed 3-colour images using the 131\,\AA, 193\,\AA, and 171\,\AA~channels of AIA, to which we assign the red, green, and blue colours, respectively (see e.g.~Fig.~\ref{Fig:Hot_loop}j-m). Therefore, features emitting in \ion{Fe}{VIII} in 131\,\AA~and in \ion{Fe}{IX} in 171\,\AA~appear to be magenta-coloured, while the purely hot \ion{Fe}{XXI} plasma appears in distinctly red colour. To increase contrast, the observations were processed using the multi-Gaussian normalisation (MGN) of \citet{Morgan14}.
\subsubsection{Solar Orbiter/EUI/FSI}
\label{Sect:FSI}
The event was also observed in the 304\,\AA~and 174\,\AA~EUV passbands of the Full Sun Imager, which is a part of the Extreme Ultraviolet Imager onboard the Solar Orbiter spacecraft \citep[Solar Orbiter/EUI/FSI,][]{Rochus20,Auchere23}. Solar Orbiter has passed perihelion on 2022 March 26 
and reached quadrature with Earth few days later on 2022 March 29 
\citep[][]{Berghmans23}. 
Table~\ref{Table2} and Fig.~\ref{Fig:Spacecrafts} give details about the spacecraft position and difference in the light travel time due to its trajectory.
The size of the Sun as observed by Solar Orbiter/EUI/FSI also varies with time. On 2 April 2022, the solar diameter as seen by Solar Orbiter was about 2695$\arcsec$ and the AR complex was located in its eastern solar hemisphere. The pixel size of Solar Orbiter/EUI/FSI observations is 4$\farcs$44 and its effective spatial resolution is about 9$\arcsec$.

The 304\,\AA~passband of Solar Orbiter/EUI/FSI is dominated by the \ion{He}{II} emission just as the SDO/AIA 304\,\AA~one, while the 174\,\AA~passbands contains emission lines of \ion{Fe}{IX} and \ion{Fe}{X}, but is dominated by \ion{Fe}{X} formed around 1\,MK \citep[cf., Figs. 1 and 3 in][]{Dudik09}. Thus, the temperature response of the 174\,\AA~passband is different with respect to AIA 171\,\AA\,
\citep[see, e.g. Figure 1 in][]{Chen2021}.
We note FSI does not observe 304\,\AA~and 174\,\AA~simultaneously, as the passband observed is selected by a filter wheel. In our dataset, the cadence of the 174\,\AA~data is 10\,minutes, while for 304\,\AA~it is 30 minutes. While this cadence does not permit to study the short dynamical processes during the eruption, the FSI data are still useful for identifying changes in the flaring active region, studying the relationship of the Solar Orbiter/STIX hard X-ray sources with the magnetic environment of the flare (Sect.~\ref{Sect:X_obs}), as well as identification of co-temporal structures in corresponding AIA observations.
\subsubsection{STEREO-A/EUVI}
The event was also observed in EUV by the EUVI instrument onboard STEREO-A \citep{Howard08}. The STEREO-A is one of two identical spacecrafts launched in 2006 on orbits similar to that of Earth, with STEREO-A having slightly smaller semi-major axis than Earth, thus orbiting the Sun ahead of Earth. 
On 2 April 2022, it observed the filament eruption as a limb event (see Table~\ref{Table2} and Fig.~\ref{Fig:Spacecrafts}), and we use the STEREO-A data to supplement the SDO/AIA and Solar Orbiter/EUI/FSI observations.

The EUVI images the Sun at four EUV wavelengths of 171\,\AA, 195\,\AA, 284\,\AA, and 304\,\AA~with a spatial scale of 1$\farcs$6. Its cadence varies over time and is also dependent on the passband. Here, we use the data from the 171\,\AA~and 304\,\AA~passbands which are also observed by SDO/AIA. During the event, the cadence of both the 171\,\AA~and 304\,\AA~was several minutes. The wavelength response of both passbands are similar to that of AIA; that is, they are dominated by \ion{Fe}{ix} and \ion{He}{ii}, respectively \citep{Howard08}.
\subsubsection{Hinode/XRT}

Additional imaging observations of the event were made by the X-ray Telescope onboard the Hinode spacecraft \citep{Golub07}. Hinode orbits the Earth on a polar orbit. The XRT instrument observed the event in several passbands at resolution of about 2$\arcsec$. The cadence of XRT is variable, as the event is first observed in Be-thin (at cadence of about 90\,s) and later, from about 13:23\,UT, also in the Be-med and Be-thick filters. 
We employ the XRT Be-thin filter as it detects presence of hot plasma during the early phases of the filament eruption. The passband of the Be-thin filter 
contains contributions from lines and continuum formed at about 7--17\,\AA. 
This spectral region is relatively crowded \citep[e.g.][]{ODwyer14,Dudik19xray}, with multiple lines of ions such as \ion{Fe}{xvii}, \ion{Fe}{xviii}, \ion{Mg}{xi}, \ion{Ne}{xi}, as well as continuum formed at broad ranges of temperatures.

We employ the standard \texttt{xrt\_prep} data processing to convert the data to level 1 data products, but without cosmetic correction of the contamination spots. The XRT Be-thin data are co-aligned manually with AIA 94\,\AA~passband, which is morphologically the most similar.
\begin{figure}[!ht]
    \centering
     \resizebox{\hsize}{!}{\includegraphics{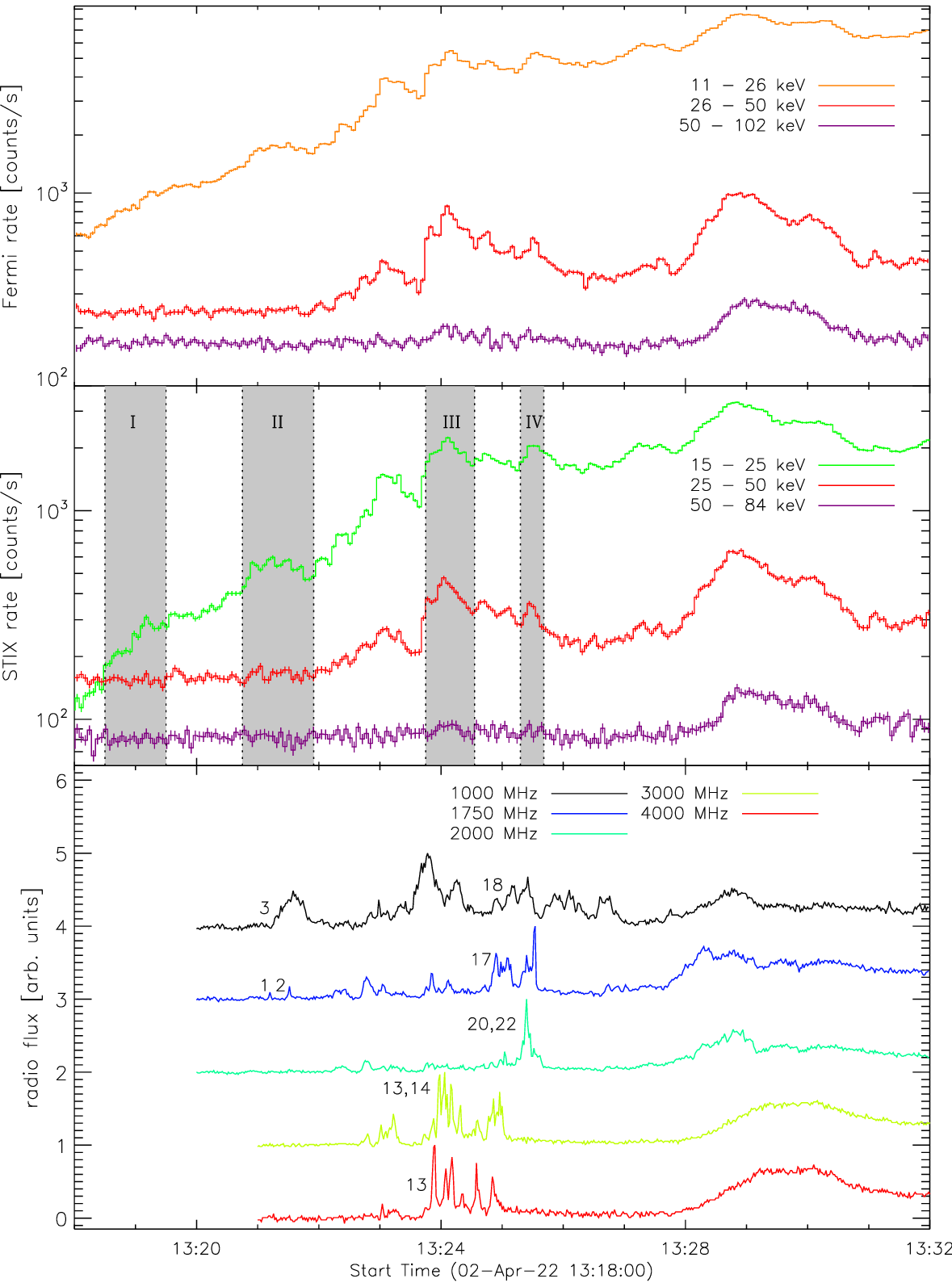}}
    \caption{Fermi/GBM detector 4 (top), Solar Orbiter/STIX (middle) count rates, and Ond\v{r}ejov radio flux at selected frequencies (bottom) as function of time during 2 April 2022 eruption. 
    Selected radio bursts are numbered (Figures~\ref{Fig:radio_spdb}-\ref{Fig:radio_1324_1325} and Table~\ref{Table1}). Four grey time intervals, I-IV (Table~\ref{tab:stix_time_intervals}), denote the STIX X-ray time intervals analysed in detail. All data are presented with respect to the time of the event detection on Earth orbit. Missing radio data at the beginning of the event were not archived because no radio bursts were observed at this time.}
    \label{Fig:HXR_evolution}
\end{figure}
\subsection{Hard X-ray observations}
\label{Sect:X_obs}
In hard X-rays (HXR) the event is observed by both Fermi/GBM and Solar Orbiter/STIX instruments. The Gamma-ray Burst Monitor (GBM) on board Fermi \citep{Meegan09} detects transient sources and is sensitive to X-rays and gamma rays with energies between 8 keV and 40 MeV. Solar flare data are provided via Fermi GBM Solar Flare Catalog. 

The Spectrometer Telescope for Imaging X-rays \citep[STIX,][]{Krucker20,Xiao2023} onboard Solar Orbiter is an X-ray imaging spectrometer covering the energy range from 4 to 150 keV. It observes HXR emission of solar flares and provides their HXR spectra and images using a Fourier-transform imaging technique.  
Although both instruments detect the flare from two different vantage points, their lightcurves are similar, see Fig.~\ref{Fig:HXR_evolution}. Thus, Solar Orbiter/STIX data are used to compare in detail with radio and EUV observations.  Specifically, we analyzed the science data file with unique identifier (UID) 2204020676, which contains pixel data at 1~s time resolution in the 4--150~keV energy range, and spectrogram data with UID 2204029033.

We selected four main STIX time intervals for a detailed imaging and spectral analysis, see Fig.~\ref{Fig:HXR_evolution} and Table~\ref{tab:stix_time_intervals}. The time intervals I and II cover the very start of the impulsive phase with the first peaks in HXR emission above 15~keV. The first significant peak above 25~keV is contained within the time interval~III, the time interval~IV corresponds to a short spike prominent during the decay of the 25--50~keV emission.

\begin{table}
\caption{STIX time intervals}
    \centering
    \begin{tabular}{ccc}
    \hline
    \hline
    No. & Time at Solar Orbiter  & Time on Earth orbit \\
     & (UT, onboard) & (UT) \\
     \hline
      I    & 13:13:09 -- 13:14:09& 13:18:30 -- 13:19:30\\
      II   & 13:15:24 -- 13:16:34& 13:20:45 -- 13:21:55\\
      III  & 13:18:24 -- 13:19:12& 13:23:45 -- 13:24:33\\
      IV & 13:19:57 -- 13:20:20& 13:25:18 -- 13:25:41\\
      \hline
    \end{tabular}
    \label{tab:stix_time_intervals}
\end{table}
\begin{figure*}[t]
        \centering
        \includegraphics[height=6cm,viewport=0 0 288 288,clip]{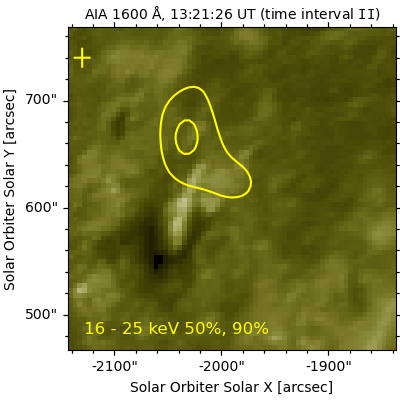}
        \includegraphics[height=6cm,viewport=16 0 288 288,clip]{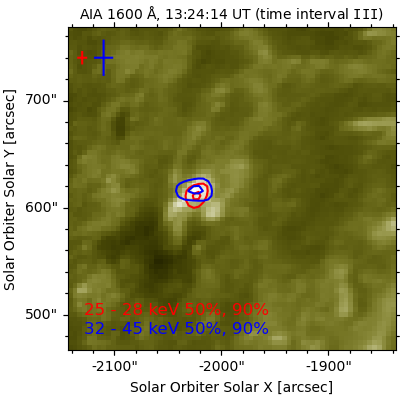}
        \includegraphics[height=6cm,viewport=16 0 288 288,clip]{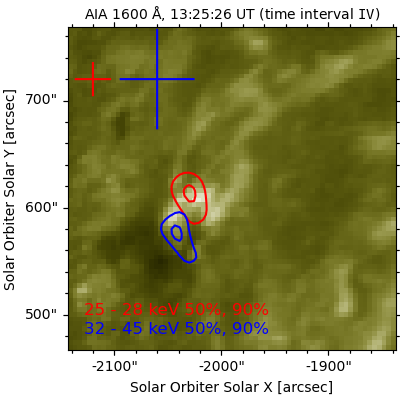}
      \caption{STIX sources in 16--25~keV (yellow), 25--28~keV (red), and 32--45~keV (blue) ranges in the STIX time intervals II-IV, overlayed on 
      AIA 1600~\AA~images reprojected into Solar Orbiter view. Contours indicate 50 and 90~\% levels of the peak flux in the STIX images, resp. Error bars denote uncertainties of the STIX image centre of mass within the 50\% contour. FOV and intensity range are the same for all AIA images.
      }
      \label{Fig:stix_1318}
\end{figure*}
\begin{figure*}[ht]
    \centering
    \includegraphics[width=8.8cm]{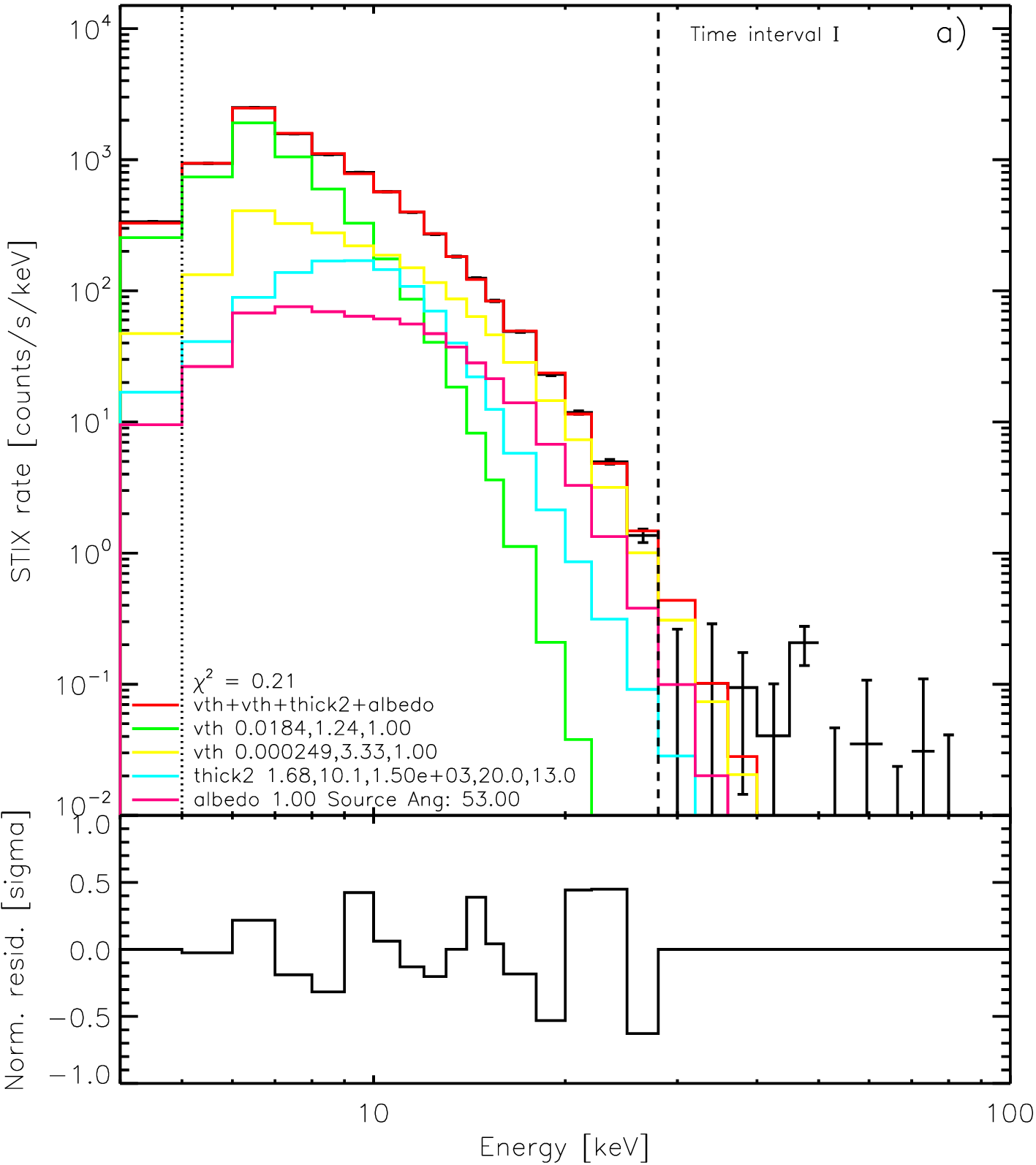}
    \includegraphics[width=8.8cm]{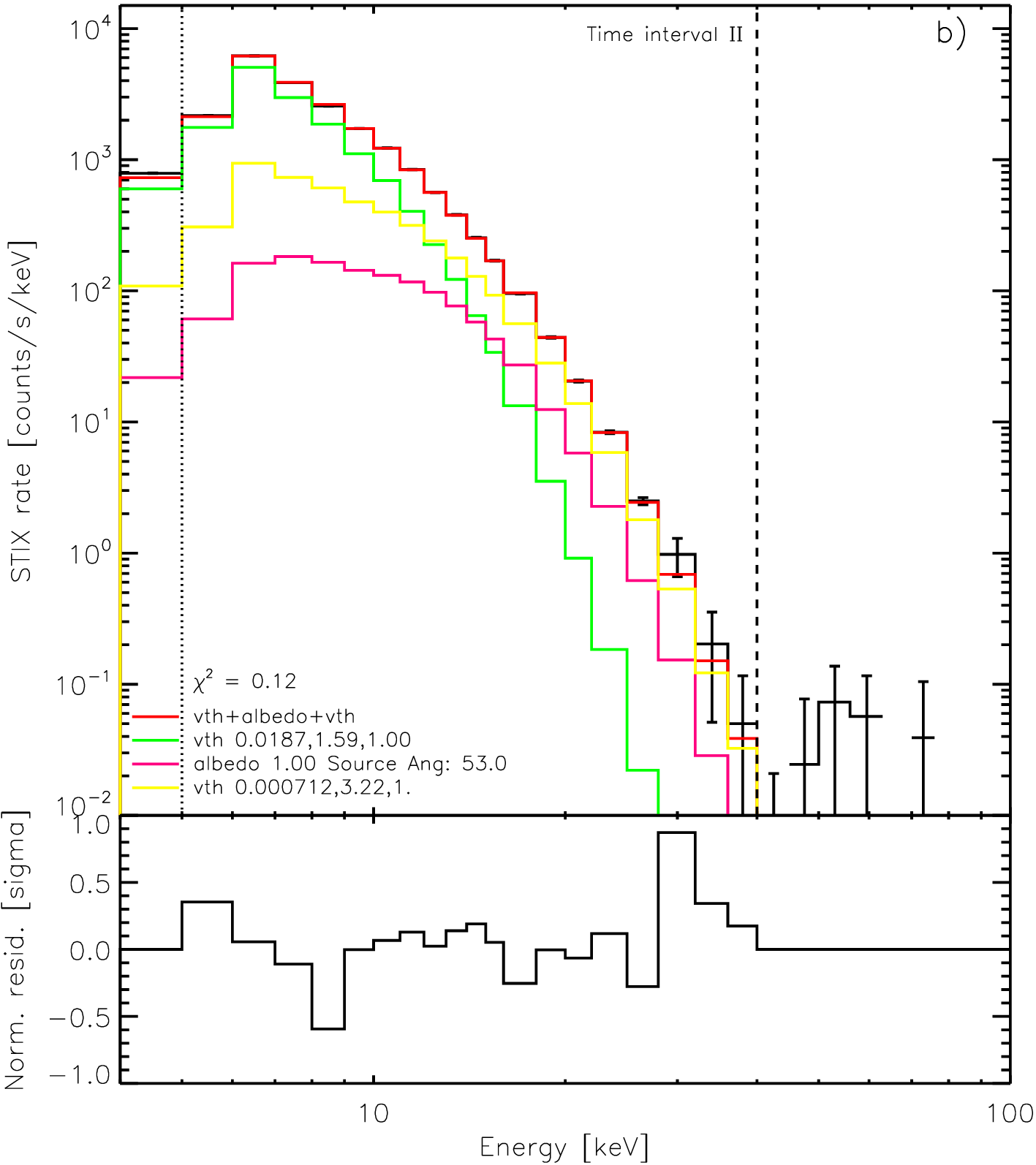}\\
    \includegraphics[width=8.8cm]{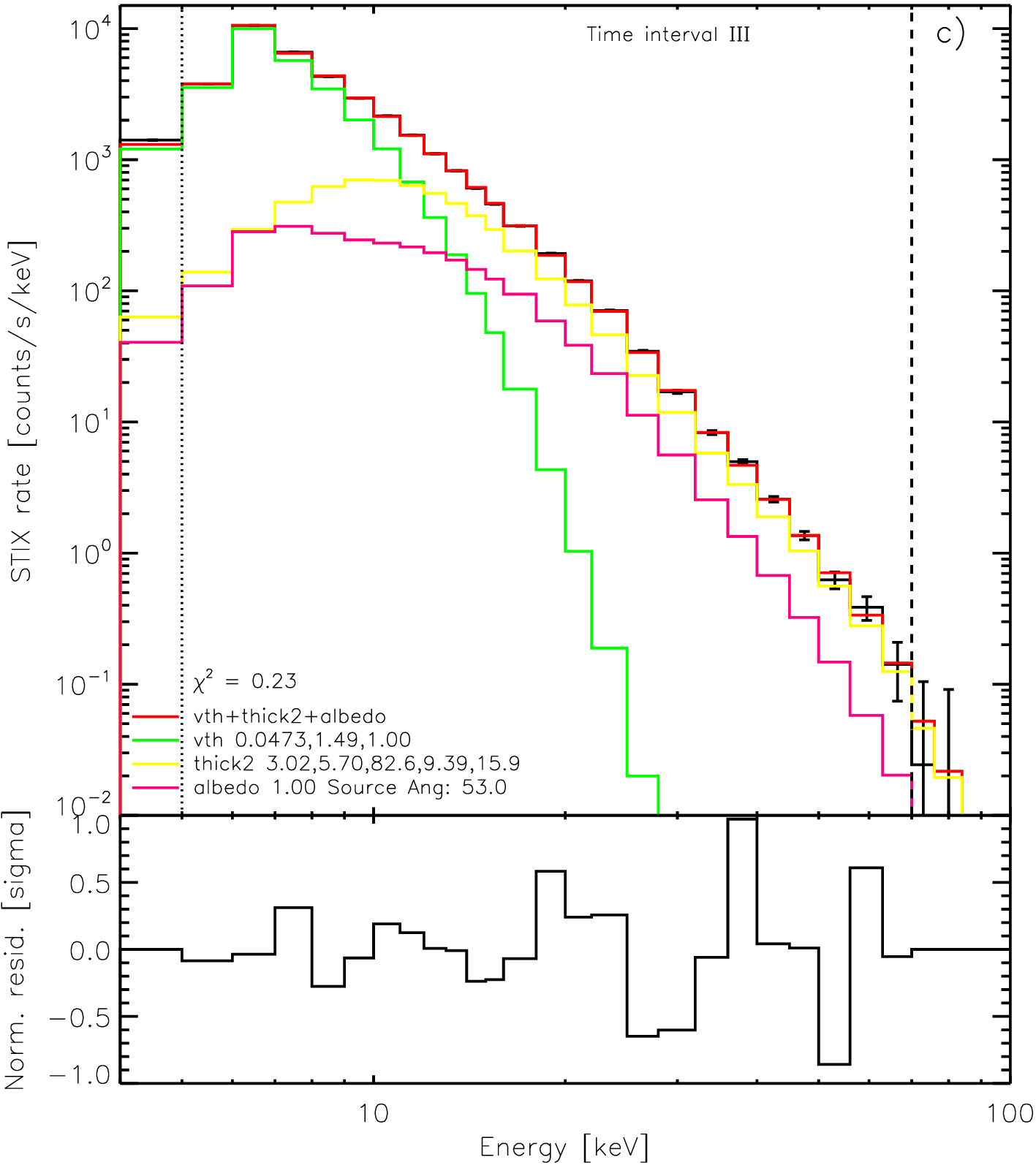}
     \includegraphics[width=8.8cm]{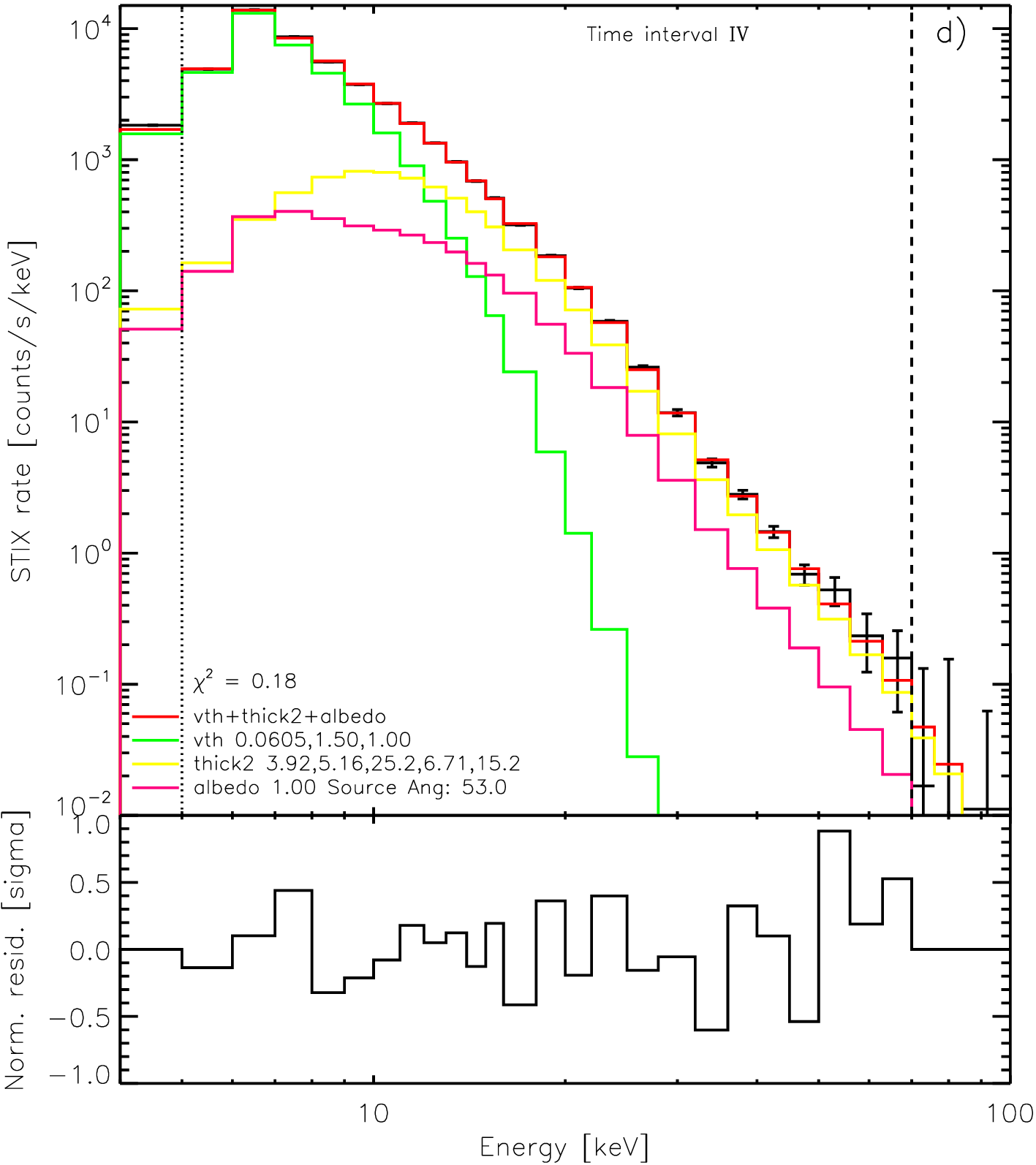}
    \caption{STIX spatially integrated count rate spectra and fits in time intervals I--IV (a--d). Vertical error bars represent uncertainties.  
    Dotted and dashed lines delimit the lower and upper energies used in the spectral fitting. The values of the reduced $\chi^2$ and the fitted parameters are indicated in the bottom left corners. For the thermal component \texttt{vth} these are emission measure [$10^{49}\mbox{cm}^{-3}$], temperature [keV], and relative abundance; for the non-thermal component \texttt{thick2} these are total electron flux [$10^{35}\mbox{electrons s}^{-1}$], spectral index above the break energy, break energy [keV], spectral index below the break energy, and the low-energy cutoff [keV], resp. Panels below the fits show normalised residuals, i.e. scaled by the corresponding uncertainties, as a function of energy.}
    \label{Fig:STIX_spectra}
\end{figure*}
STIX sources were reconstructed using MEM\_GE algorithm \citep{Massa20} implemented for the STIX data in the Solar SoftWare\footnote{\url{https://www.lmsal.com/solarsoft/}} (SSW). The MEM\_GE algorithm is based on a maximum entropy approach and uses visibilities, i.e. calibrated Fourier transforms of the incoming photon flux detected at specific spatial frequencies that are defined by the parameters characterising the STIX sub-collimators. 
Amplitude and phase of the complex visibility values are related to contrast and location of the Moir\'{e} pattern created by a pair of grids in a sub-collimator.
Measurements of the patterns provided by the pixels of the detectors located behind the grids allow determining the values of the Fourier components (for more details on the STIX imaging technique see \citet{Massa2023}).
As the visibility amplitudes decrease with increasing source size, we considered the visibilities measured by the coarse-resolution detector only (which are labelled with 6 through 10) for the reconstruction of extended sources.
For reconstructing compact sources, we included fine-resolution sub-collimators (down to label 3).
Reliability of reconstructed STIX images was tested by comparing with other reconstruction algorithms such as CLEAN~\citep{Massa2023,Hogbom1974}.

Knowledge of the STIX pointing is obtained from the STIX Aspect system \citep{Warmuth2020}, which usually provides an absolute accuracy of order 10\arcsec. However, the aspect solution that is used for imaging within STIX SSWIDL was not accurate enough for this particular event.
Therefore, we followed the same approach as previous studies \citep[e.g.][]{Purkhart23} and co-aligned STIX non-thermal sources with a relevant AIA 1600~\AA\, image reprojected to the Solar Orbiter vantage point (neglecting possible different altitudes of AIA 1600~\AA\, structures above reference solar surface). By doing so, we assumed that the localised emission seen in that AIA filter is the UV chromospheric/transition-region counterpart of the non-thermal STIX sources. Although the spectral content of the AIA UV filters does not originate only in chromosphere and below \citep{Simoes2019}, localised and  prominent brightenings related to flare footpoints are known to be present also in UV, EUV, and X-ray emission \citep[e.g.][]{2011SSRv..159...19F}, indicating strongly multi-thermal plasma.
The EUI/FSI images were not used for co-alignment of STIX with EUV emission due to limited FSI 304\,\AA~cadence and complex EUV structures seen in the 174\,\AA~channel (Fig.~\ref{Fig:euv_1325UT}a). We determined the spatial offset of STIX images in the time interval III by shifting STIX non-thermal sources at energies above 25 keV to the position of the brightest AIA UV emission, see Fig.~\ref{Fig:stix_1318} (middle). The same spatial offset was then applied to all other STIX images.

STIX images reconstructed at highest energies detected during the time intervals II-IV are displayed in Fig.~\ref{Fig:stix_1318}. The energy bin 28--32~keV was excluded to omit a significant signal from the onboard Ba~133 calibration source. Uncertainties of the source positions are indicated in the upper corners of each panel and correspond to the uncertainty of the image centre of mass within the 50\% contours. We determined those uncertainties as follows: STIX data (visibilities) were randomly varied within a normal distribution of their corresponding standard deviation. The set of modified data was processed by MEM\_GE algorithm to obtain a set of reconstructed sources. Then, the positions of the centre of mass within the 50\% contour were collected. Finally, the standard deviations of those positions in x- and y- directions represent the uncertainties of source position due to the uncertainties in the STIX data. The similar approach is implemented in the SSW tree for the FWDFIT\_PSO algorithm \citep{Volpara2022}. Generally, the uncertainties are larger for weaker sources and at higher energies where data uncertainties are larger, see Fig.~\ref{Fig:STIX_spectra}. Also, we point out that these uncertainties do not represent the uncertainty on the absolute position with respect to the AIA or any other imaging data.

STIX spatially integrated spectra corresponding to the time intervals I-IV together with their fitting results are displayed in Fig.~\ref{Fig:STIX_spectra}. Spectral fitting was done within the SSWIDL OSPEX\footnote{\url{https://hesperia.gsfc.nasa.gov/ssw/packages/spex/doc/ospex_explanation.htm}} environment using the forward-fitting approach. We used STIX IDL software version 0.5, systematic uncertainties of 5\%, and OSPEX functions \texttt{vth}, \texttt{thick2}, and \texttt{albedo} to describe iso-thermal, non-thermal power-law in the thick-target approximation \citep{1971SoPh...18..489B}, and albedo components, respectively. The albedo component takes into account contribution of HXR photons scattered in the photosphere \citep{Kontar2006}. STIX flare viewing angle of \ang{53} 
and an isotropic primary source of HXR photons were assumed in all fits. The spectrum of the time interval III is consistent with a thermal component plus a non-thermal component due to a power-law electron beam which dominates the photon spectrum above $\sim 15$~keV. Therefore, the reconstructed HXR sources above 25~keV for this interval could be considered as chromospheric and used for the co-alignment with the AIA 1600~\AA\, emission, see Fig.~\ref{Fig:stix_1318} (middle).
\section{Eruption in radio, EUV, and X-rays}
\label{Sect:Evolution}
\subsection{Evolution of the filament eruption}
\label{Sect:Evolution_EUV}
\begin{figure*}
    \begin{center}
      \includegraphics[width=5.1cm,viewport = 10 35 305 255,clip]{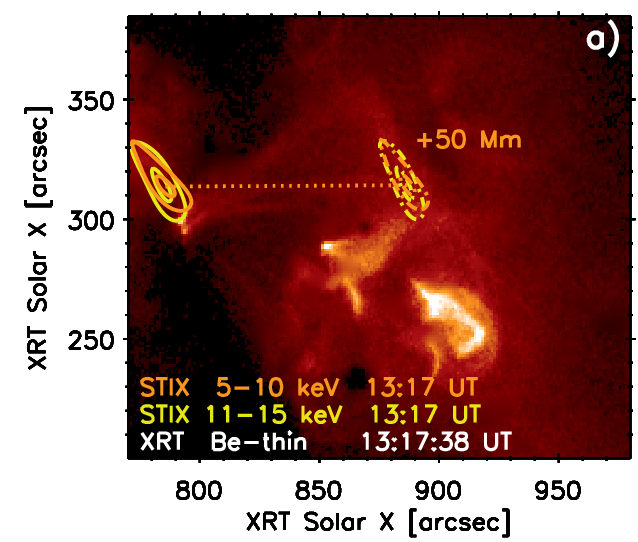} 
      \includegraphics[width=4.23cm,viewport = 60 35 305 255,clip]
      {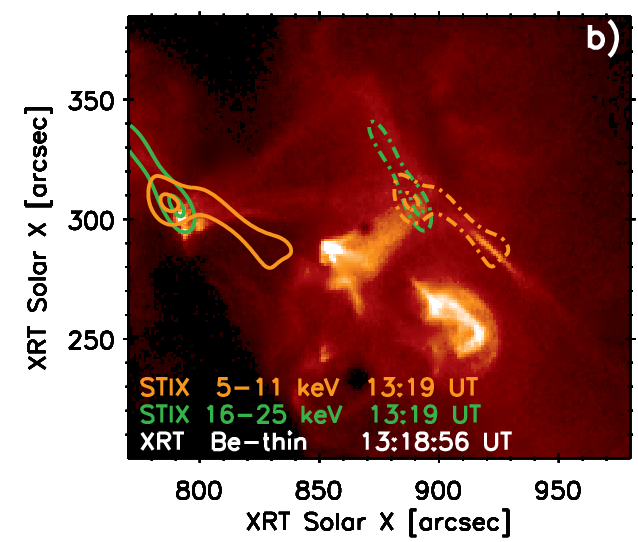}
      \includegraphics[width=4.23cm,viewport = 60 35 305 255,clip]{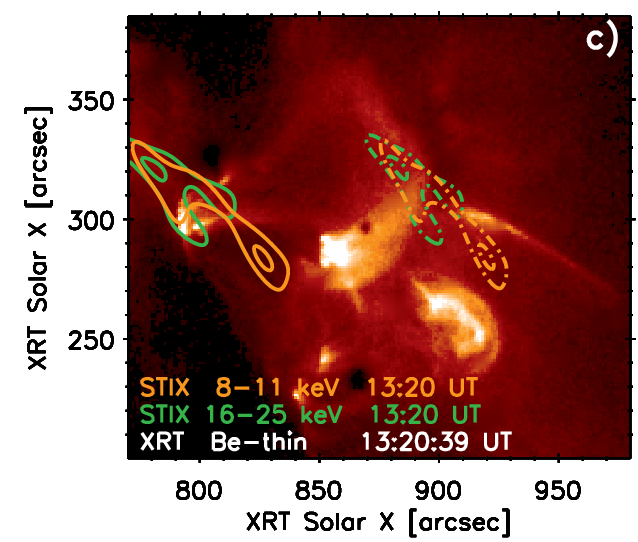} 
      \includegraphics[width=4.23cm,viewport = 60 35 305 255,clip]
      {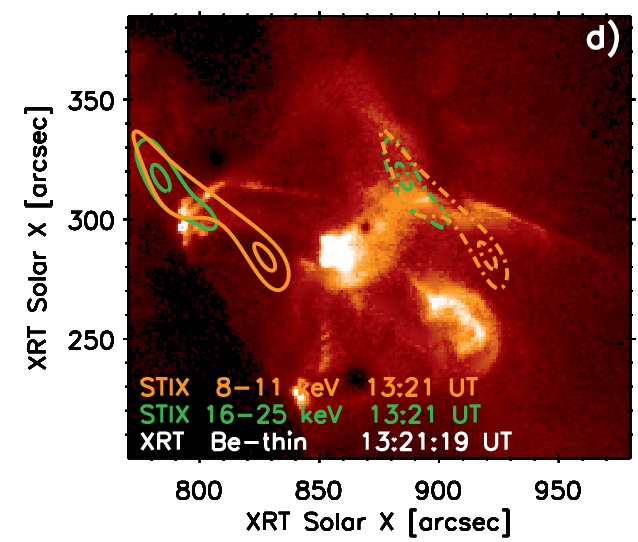}

      \includegraphics[width=5.10cm,viewport = 10 35 305 255,clip]
      {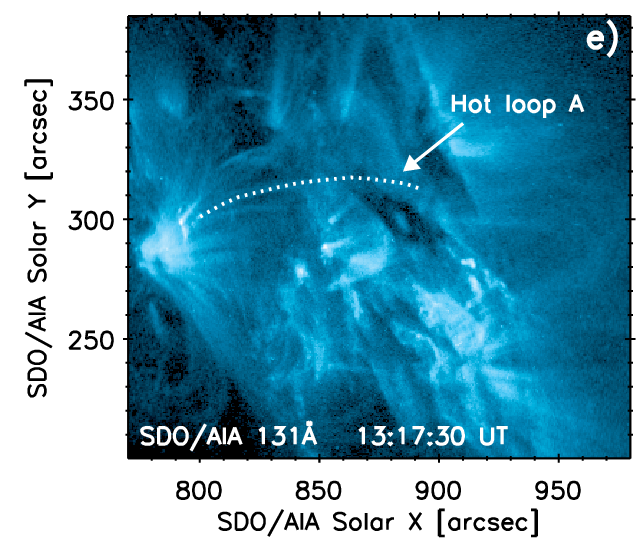}
      \includegraphics[width=4.23cm,viewport = 60 35 305 255,clip]
      {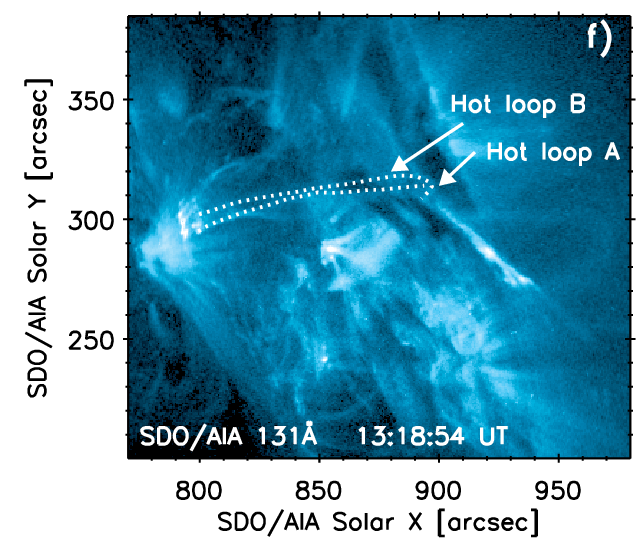}
      \includegraphics[width=4.23cm,viewport = 60 35 305 255,clip]{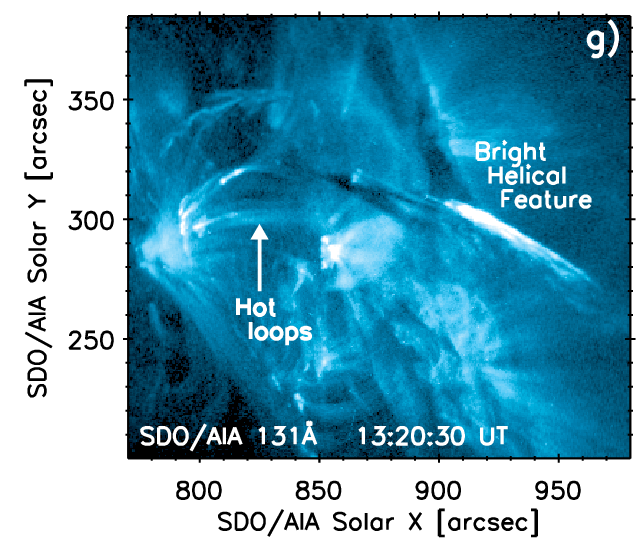} 
      \includegraphics[width=4.23cm,viewport = 60 35 305 255,clip]
      {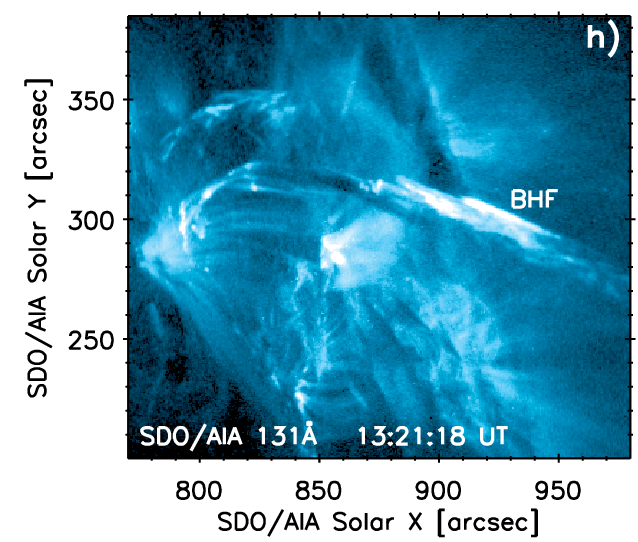}
      \includegraphics[width=5.10cm,viewport = 10 2 305 255,clip]{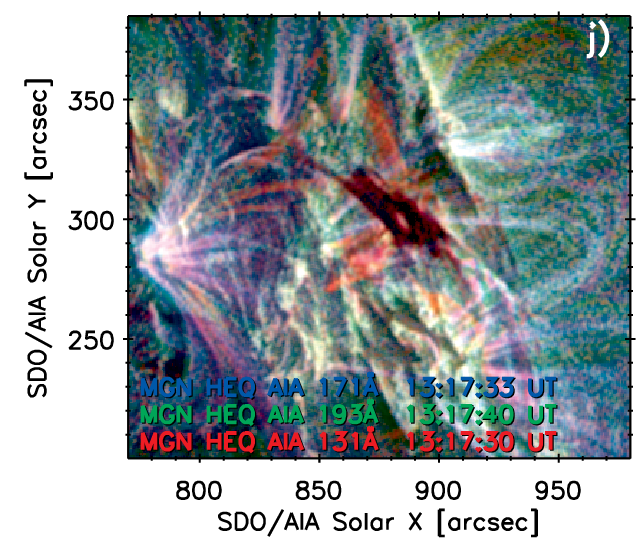} 
      \includegraphics[width=4.23cm,viewport = 60 2 305 255,clip]
      {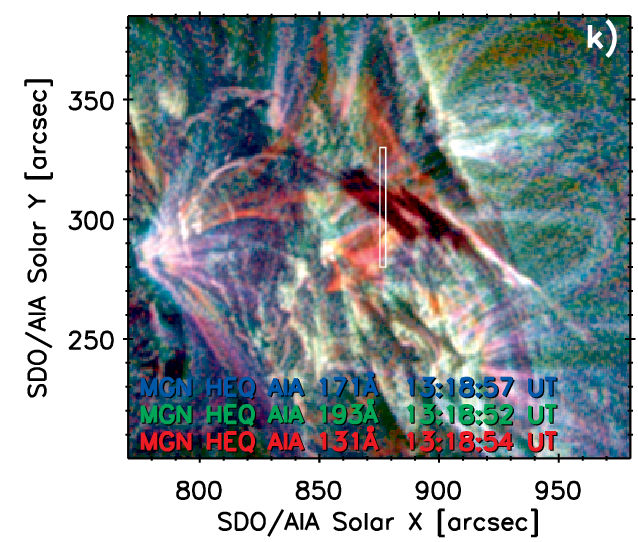}
      \includegraphics[width=4.23cm,viewport = 60 2 305 255,clip]
      {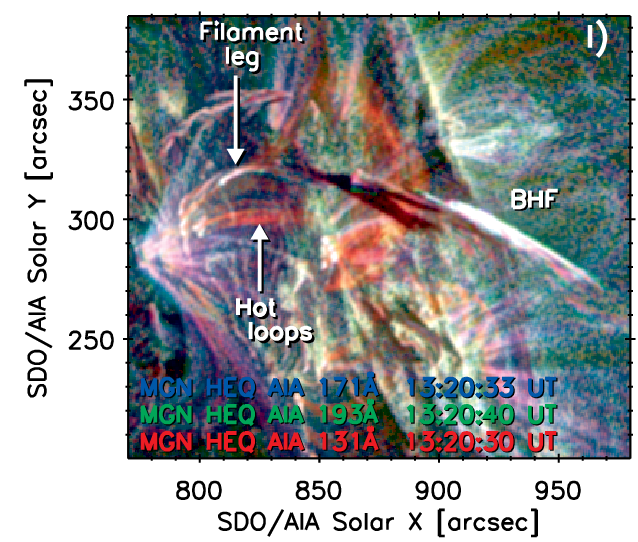} 
      \includegraphics[width=4.23cm,viewport = 60 2 305 255,clip]{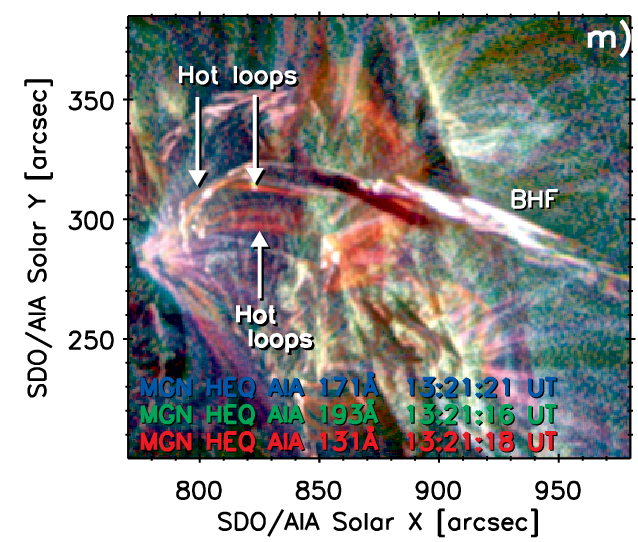} 
      \end{center}
      \caption{Filament interaction with hot arcade above. Dotted white curves denote hot loops A and B seen in XRT and AIA~131\,\AA. 
      The bright helical feature (BHF) develops in their vicinity. Contours indicate 50 and 90~\% levels of the 5--10 keV and 11--15 keV STIX sources reprojected assuming 0 and 50\,Mm height (full and dash-dotted lines, resp.). 
      Bottom row: MGN composites of 131\,\AA~(red), 193\,\AA~(green), and 171\,\AA~(blue).
      (k) slit position for the time-distance plot (Fig.~\ref{Fig:Hot_loop_304}a) in white.
      Animation of the MGN composites with the slit position during 13:10--13:30~UT is available online. 
      \label{Fig:Hot_loop}}
    \begin{center}
      \includegraphics[width=8.41cm,viewport = 0 37 550 280,clip]
      {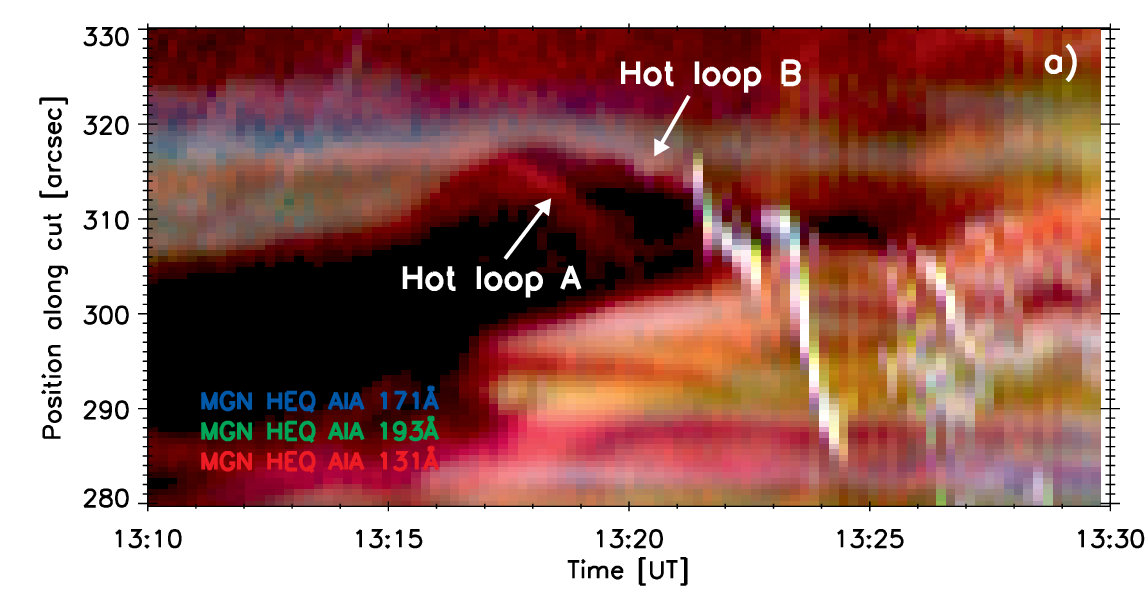}
      \includegraphics[width=5.10cm,viewport = 10 35 305 255,clip]
      {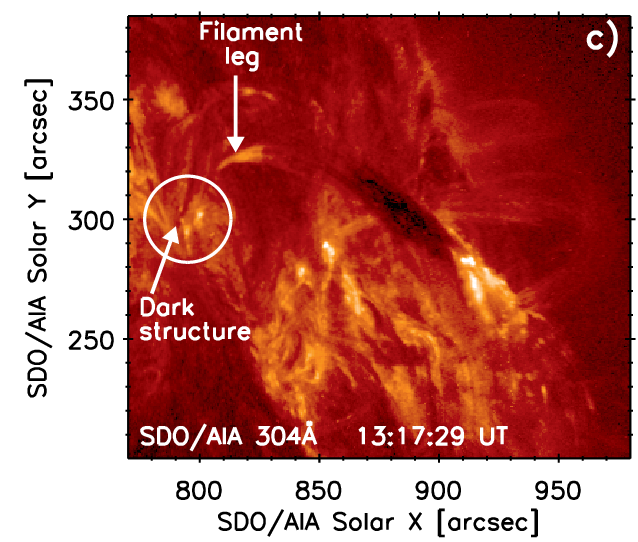}
      \includegraphics[width=4.23cm,viewport = 60 35 305 255,clip]
      {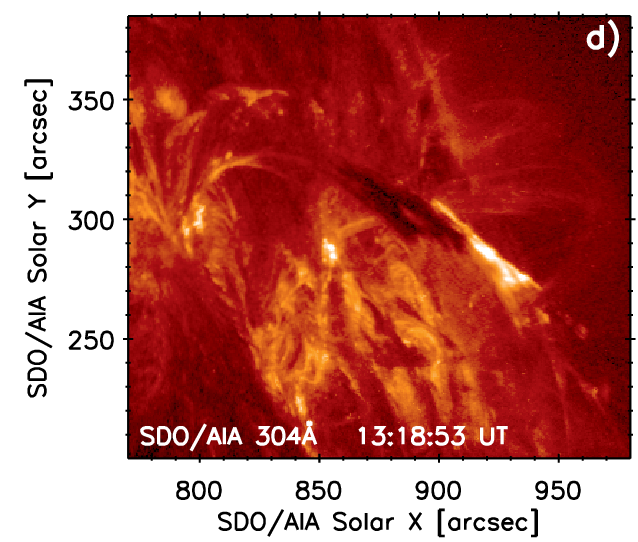}
      \includegraphics[width=8.41cm,viewport = 0  0 550 280,clip]{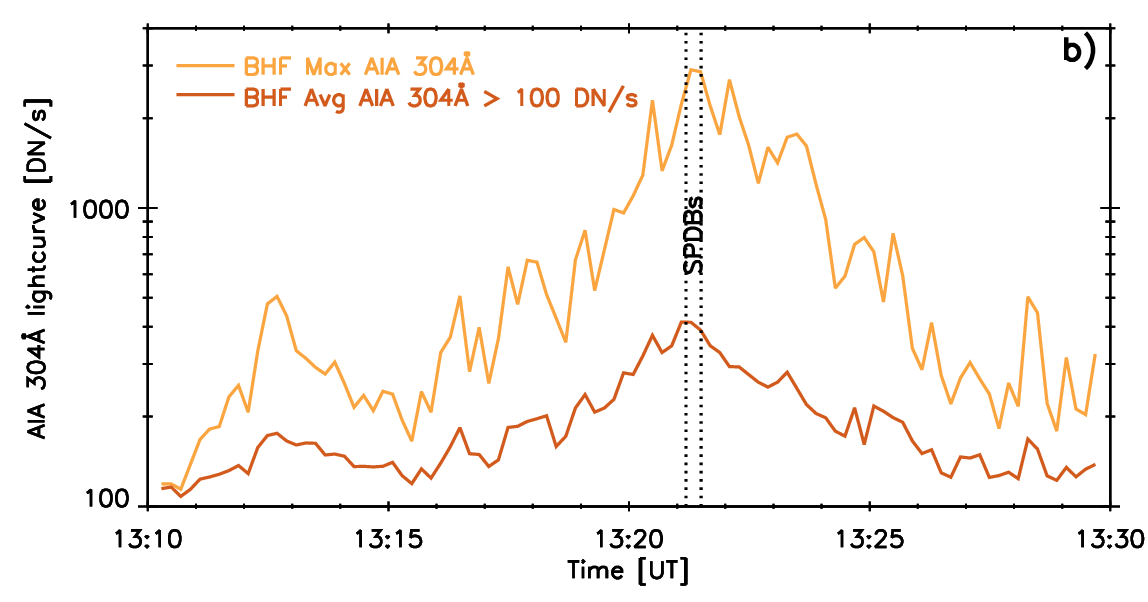}
      \includegraphics[width=5.10cm,viewport = 10  2 305 255,clip]{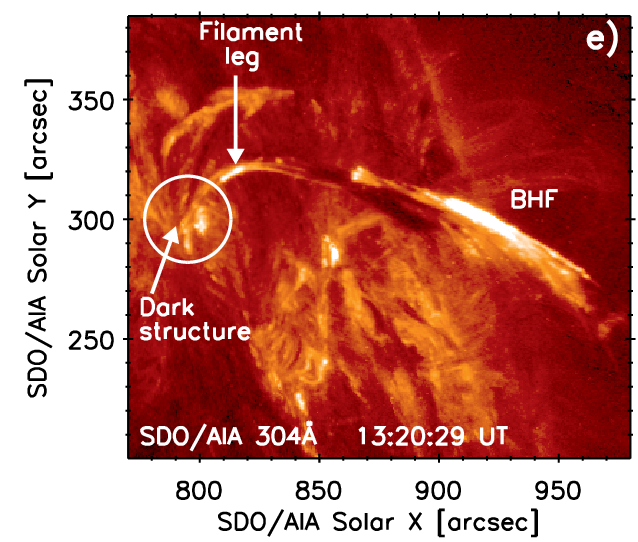} 
      \includegraphics[width=4.23cm,viewport = 60  2 305 255,clip]{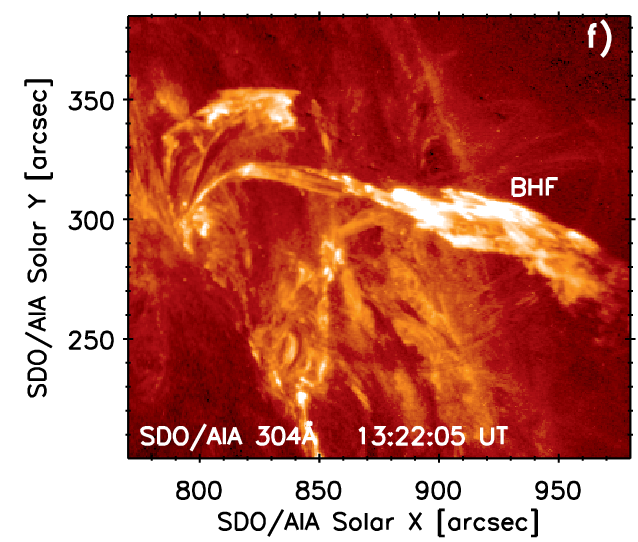} 
    \end{center}
\caption{Filament interaction with hot arcade above. (a) MGN time-distance plot along the slit (Fig.~\ref{Fig:Hot_loop}k), (b) AIA 304\,\AA~lightcurves for the maximum and average brightness within BHF. 
    Vertical dotted lines mark the time of occurrence of the SPDBs (Table~\ref{Table1}). (c-f) BHF formation and movement of the filament leg with respect to the dark structure (white circle) in AIA 304~\AA\ at similar times as in Fig.~\ref{Fig:Hot_loop}.\label{Fig:Hot_loop_304}}
\end{figure*}

After about 13:05\,UT, the eruption starts 
accompanied by gradual lifting-off of the southern leg of the filament as the filament erupts (see the animation accompanying Fig.~\ref{Fig:aia_overview}). 
The northern leg of the filament, located within the AR complex (Fig.~\ref{Fig:aia_overview}c, $[X,Y]$\,=\,$[800$\arcsec$, 300$\arcsec$]$),  shows  the most interesting evolution. Here, the rising filament interacts with the overlying arcade of hot loops (Fig.~\ref{Fig:Hot_loop}). 

During the interaction the rising part of the filament  
undergoes significant untwisting motions 
accompanied by a strong brightening observed in EUV 
(Figs.~\ref{Fig:aia_overview}b,~\ref{Fig:Hot_loop}g,~\ref{Fig:Hot_loop_304}e, f). We denote it as a "Bright Helical Feature" (BHF).
This BHF is seen as a bright structure not only in 304\,\AA, but also in 131\,\AA, 193\,\AA, and 171\,\AA\,
and that results in a distinctly white feature on the colour-composite MGN figures (Fig.~\ref{Fig:Hot_loop}l, m).

The untwisting was studied also by \citet{Janvier23} (Figs.~14, 15, and animation therein), which indicated large positive and negative LOS velocities within BHF at 13:21:37~UT. The length of one helical turn (twist angle 2$\pi$) within this BHF is much less than the length of the erupting filament at this time (see the animation accompanying Fig.~\ref{Fig:aia_overview}, 13:21-13:23\,UT). The remnant of the darkest part of the filament, visible as a dark thread underneath this BHF, wraps around the filament leg (Figs.~\ref{Fig:Hot_loop}l and~\ref{Fig:Hot_loop_304}e). 

The untwisting motions together with the expansion lead to appearance of
a series of blobs, most of which move outward.
(Fig.~\ref{Fig:aia_overview}d and accompanying animation, Fig.~\ref{Fig:euv_1325UT}c, e). 
At about 13:25 UT they form a quasi-circular structure.
By 13:30\,UT only remnants of the northern filament leg are visible (Fig.~\ref{Fig:aia_overview}e). 
Magnetic rope and filament eruptions are often related to the presence of a pre-eruption sigmoid \citep[e. g.][]{Fan2015}. In this event \citet{Janvier23} reported on a twisted pre-existing filament detected in Solar Orbiter/EUI/FSI 304\,\AA\, filter hours before the flare (Fig. 4d therein). We did not detected additional sigmoid or hot channel structure using SDO.
\subsection{Hot plasma in the vicinity of the filament}
\label{Sect:Hot_plasma}

The filament eruption is accompanied by appearance of multiple systems of flare loops (Fig.~\ref{Fig:aia_overview}f--h) and is located close to the western limb of the Sun (SDO vantage point) seen in severe projection. \citet{Janvier23} provided an interpretation of the origin of several systems of closed flare loops in terms of a few individual reconnections: the primary one, standard flare reconnection producing the closed flare loops below the erupting flux rope as well as the new flux rope field lines wrapping around the flux rope; followed by the secondary and tertiary reconnections with the magnetic flux originating in the emerging bipole (see Fig.~19 therein). 

In the following, we do not focus on these systems of flare loops. Instead, we describe the presence of additional hot loops, not described by \citet{Janvier23}, in the close vicinity of the filament, or within it; as well as their relation to the HXR sources at the very start of the filament rising. 
Hot loops are relatively bright in AIA 131\,\AA, having up to $\approx$~100~DN\,s$^{-1}$. They are seen also in XRT Be-thin filter, this hot emission contributes to STIX data at lower energies too. Evolution of STIX X-ray sources at energies below 25 keV and their relation to EUI/FSI 174~\AA\, images is summarised in Appendix~\ref{App1} and Fig.~\ref{fig_eui_stix_evol}.

AIA 131\,\AA~and XRT Be-thin observations show presence of hot plasma as early as 
13:09~UT (see the animation accompanying Fig. \ref{Fig:Hot_loop}), 
although the rise of the filament started even earlier 
\citep[Sect. ~\ref{Sect:Evolution_EUV} and also Fig.~14 left of][]{Janvier23}.
The evolution of the BHF and the associated hot plasma in its vicinity is shown
in Fig.~\ref{Fig:Hot_loop}. By 13:17:30\,UT (first column of Fig.~\ref{Fig:Hot_loop} -- panels a, e, and j) the system of hot loops became more prominent.  We detected a hot loop A overlying the rising filament. This hot loop is present in 131\,\AA, and well visible in the MGN colour-composite image (Fig.~\ref{Fig:Hot_loop}j). A portion of it that can be clearly traced is shown as a dotted white line overlaid in Fig.~\ref{Fig:Hot_loop}e. That the loop is closed can be discerned from the XRT Be-thin image (Fig.~\ref{Fig:Hot_loop}a), where the filament itself is not visible \citep{Heinzel08}.

At the same time, $\sim$~13:17:40~UT, STIX spectrum can be fitted  
equally well with several combinations of thermal or non-thermal components.
The thermal component dominated 
the 5--10~keV range. The 11--15~keV source is very similar to the thermal one (Figs.~\ref{Fig:Hot_loop}a and \ref{fig_eui_stix_evol}b), although spectral fitting of the spatially integrated spectrum resulted in a thick-target component dominating this range. However,
the modulation pattern is prominent namely in the coarser sub-collimators 9 and 10, similarly as for the thermal 5--10~keV source. 
The location of a coronal/thermal source when reprojected to AIA has one free parameter - the height of the source above the solar surface. In Fig.~\ref{Fig:Hot_loop}a we show both sources at zero height as well as at height of 50 Mm, where 
both sources could be associated with the top part of the hot loop system. Therefore, the 11-15~keV source is possibly of a coronal thick-target type \citep{Veronig2004}. 

As the time moves on, another hot loop, B, appears near hot loop A. The erupting filament rises and reaches the hot loops. At 13:19\,UT (second column of Fig.~\ref{Fig:Hot_loop} -- panels b, f, and k), the loop A is directly above the filament, while the BHF starts to develop along the filament directly at this location.
After $\sim$~13:18~UT both STIX sources, observed at 5--10~keV and 11--15~keV, started to split spatially.
At $\sim$~13:19~UT (STIX time interval I) the spatial splitting was already prominent, see Fig.~\ref{Fig:Hot_loop}b. 
The 5--10~keV thermal source could be co-aligned with BHF, while 
the 16--25~keV source could correspond to the top part of the hot arcade where BHF and loops A and B met.
The corresponding STIX spectrum (Fig.~\ref{Fig:STIX_spectra}a) 
was fitted well with two thermal and a thick-target component, the 16--25~keV range being dominated by super-hot thermal component of $T=3.3$~keV ($\sim 38$~MK).
We note that such super-hot sources have been previously reported e.g.~by \citet{Caspi2015,Polito18}. 

By 13:20:30\,UT, the hot loops A and B are no longer visible over the filament, and only a portion of them can be discernible, rooted at the hooked flare ribbon below ("Hot loops" in Fig.~\ref{Fig:Hot_loop}l, upward pointing arrow). 
Time-distance plot (Fig.~\ref{Fig:Hot_loop_304}a), along a slit shown in Fig.~\ref{Fig:Hot_loop}k, indicates projected movement of the hot loops A and B through the dark filament structure, later followed by BHF crossing the slit (see also movie in Fig.~\ref{Fig:Hot_loop}). 
Even though the loop B does not show up in the MGN time-distance plot as a distinct red structure, it is hot (Appendix~\ref{App:time_plot}).
The filament rise is seen as a dark edge shifting towards larger y-values starting at about 13:17 UT, whereas the movement of the upper parts of the hot loops A and B are seen as distinct curves shifting towards smaller y-values (Figs.~\ref{Fig:Hot_loop_304}a, \ref{fig_stck_layers}).

The BHF continues to brighten significantly and also propagates downwards along the filament. 
Additionally, during that time, i.e. between the STIX time intervals I and II, there was no significant change in the STIX spectral shape. Yet, the low-energy STIX thermal source, 8--11\,keV, extended again towards the top part of the hot loops  and its northern part was located at a similar place as the higher energy source, i.e at the 16--25\,keV range (Fig.\ref{Fig:Hot_loop}c).
\subsection{Features temporally associated with radio bursts}
\label{Sect:EUV_radio}

This eruption displayed a series of rare and unique bursts occurring between 13:21\, UT and 13:25\, UT, as described in Sect.~\ref{Sect:Radio_obs}. Although our radio observations lack positional information, it is still possible to identify potential corresponding source structures in EUV, provided a unique structure is observed at the time of the radio burst. Since no other eruptive events occurred on the Sun during this period, we focused on distinctive structures observed in EUV within the eruption itself.

\subsubsection{BHF at 13:21 UT}
\label{Sect:BHF}

As noted already in Sect.~\ref{Sect:Evolution_EUV}, at 13:21\,UT there is a strong brightening within the untwisting northern leg of the erupting filament. The BHF appears already at 13:19\,UT (Fig.~\ref{Fig:Hot_loop}b, f) and subsequently increases in both brightness and area, as it moves along the filament, see Fig.~\ref{Fig:Hot_loop} - third and fourth columns.  Moreover, the filament leg and its footpoint move from behind a dark inclined structure (white circle in Fig.~\ref{Fig:Hot_loop_304}c, e) and,  at 13:22~UT, the filament leg is already positioned in front of the dark structure (Fig.~\ref{Fig:Hot_loop_304}c-f).  
By that time BHF is formed and another set of hot loops appears also close to the filament leg, see the downward pointing arrows in Fig.~\ref{Fig:Hot_loop}m. This evolution can be seen in the video accompanying Fig.~\ref{Fig:aia_overview} as well.

The BHF reached peak brightness during 13:21:05--13:21:41\,UT, both in terms of maximum intensity detected in AIA 304\,\AA~(Fig.~\ref{Fig:Hot_loop_304}b, orange), 
as well as in average intensity (in areas above 100 DN\,s) within the brightening (Fig.~\ref{Fig:Hot_loop_304}b, dark orange). The spatial limits $X$\,$\in$\,$\left<870\arcsec, 1030\arcsec\right>$ and $Y$\,$\in$\,$\left<230\arcsec, 350\arcsec\right>$ were chosen to isolate the signal in BHF.

The peak brightness of BHF coincides well with the radio bursts 1 and 2 (see Table \ref{Table1} and Fig.~\ref{Fig:radio_spdb}), indicating they are likely related to the BHF. Both bursts are of the SPDB type, i.e. RS bursts of very slow frequency drift ($\sim$+97 and +77 MHz s$^{-1}$) in the GHz frequency range.

At this time (STIX time interval II), the STIX spectrum is consistent with two thermal components, see Fig.~\ref{Fig:STIX_spectra}b, although a fit with an iso-thermal and a thick-target component with similar residua can be obtained as well.
At lower energies the temperature of the thermal component is 1.6~keV ($\sim 18$~MK). The image reconstruction in the 8--11 keV range reveals a very extended source, $\sim$~100\arcsec\ in length when defined by a 50\% contour of maximum flux and viewed from the Earth vantage point, see Fig.~\ref{Fig:Hot_loop}d. 
Temperature of the second thermal component fitting the energies $\sim$16--40~keV, is $3.2$~keV $\sim 37$~MK. Thermal origin of that X-ray emission is further supported by the position of the reconstructed source in the 16--25~keV energy range. 
At the source position, no chromospheric brightening 
is present in AIA 1600~\AA\, filter, see Fig.~\ref{Fig:stix_1318} (left). Rather, there is an elongated emission, probably coming from a lower part of a loop, which is visible also in  AIA 304~\AA\ filter in Fig.~\ref{Fig:aia_overview}c at $\sim$ [800\arcsec, 300\arcsec]. 
The STIX source in the 16--25~keV is co-spatial with the NE part of the thermal source and is approximately half in size, see Fig.~\ref{Fig:Hot_loop}d. Both sources can be placed near the hot loops or their top part seen in AIA and XRT images, if a height up to $\sim$50~Mm above the solar surface is assumed, see Fig.~\ref{Fig:Hot_loop}d. Unfortunately, Solar Orbiter/EUI/FSI did not capture BHF as it occurred in between two exposures of 174~\AA\ filter (13:16 and 13:26 UT). Thus, we use the 174~\AA\ image primarily for the context location of STIX X-ray sources with respect to EUV emission seen from the same vantage point (Appendix~\ref{App1}). Nevertheless, it indicates that STIX sources partially overlie the loop system detected before the filament eruption.
\begin{figure*}[!ht]
    \begin{center}
      \includegraphics[width=6.76cm,viewport =  0  0 428 350,clip]{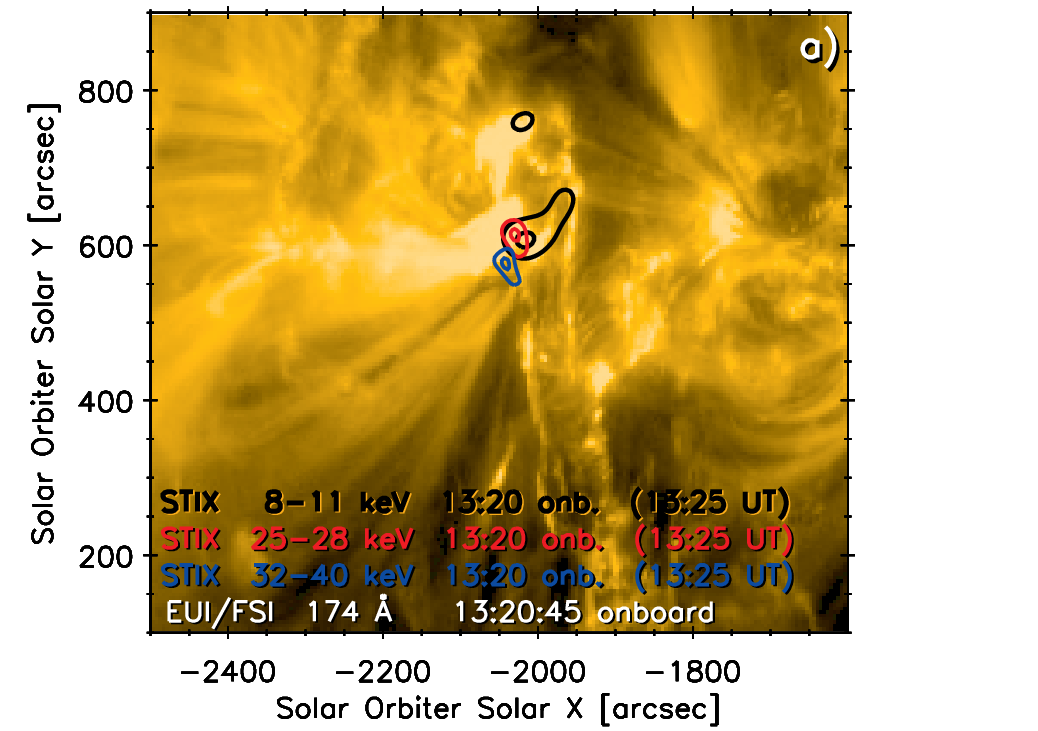} 
      \includegraphics[width=7.83cm,viewport =  0  0 496 350,clip]{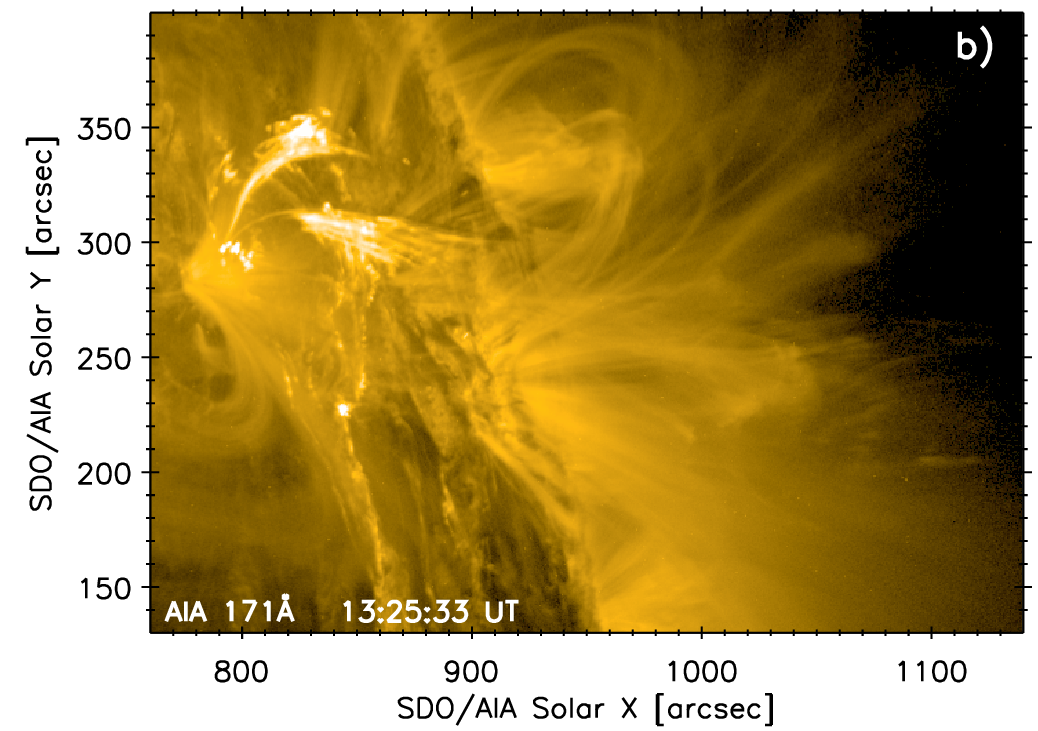}
      \includegraphics[width=7.83cm,viewport =  0  0 496 350,clip]{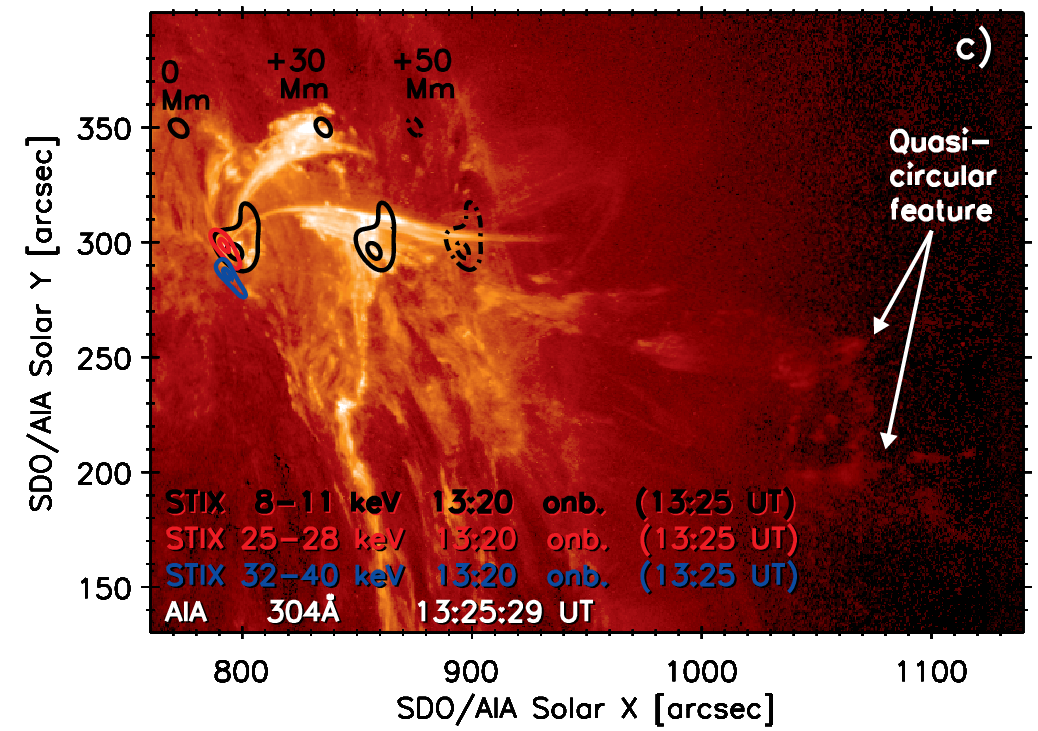}
      \includegraphics[width=6.76cm,viewport = 68  0 496 350,clip]{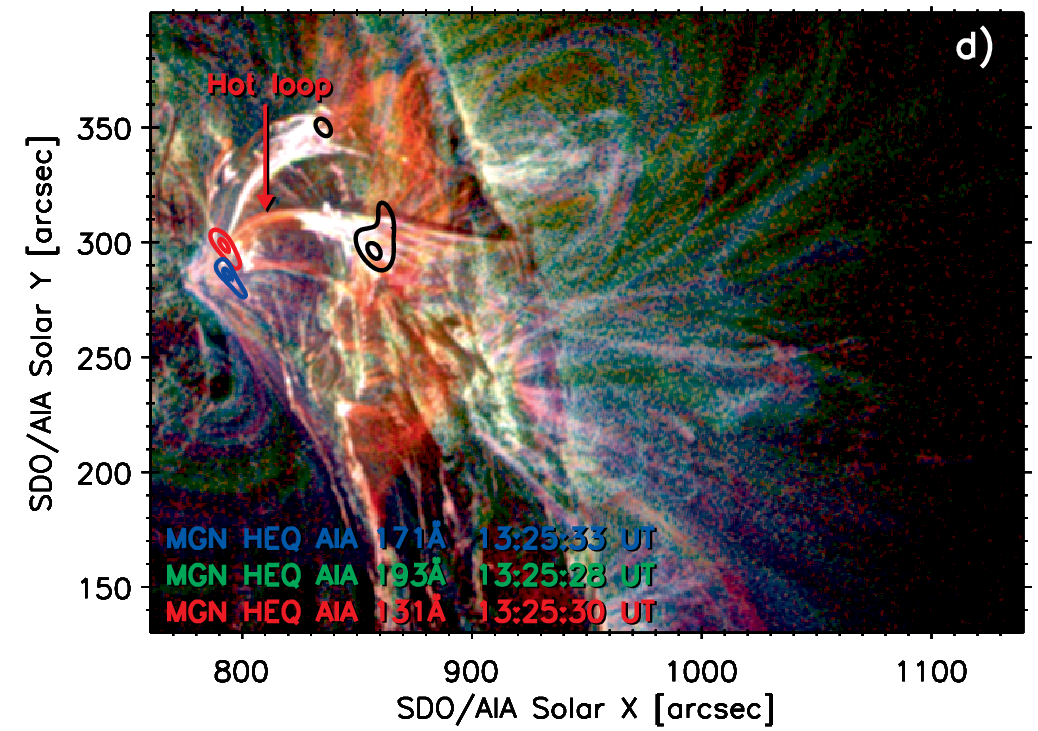}
      \includegraphics[width=7.83cm,viewport =  0  0 496 350,clip]{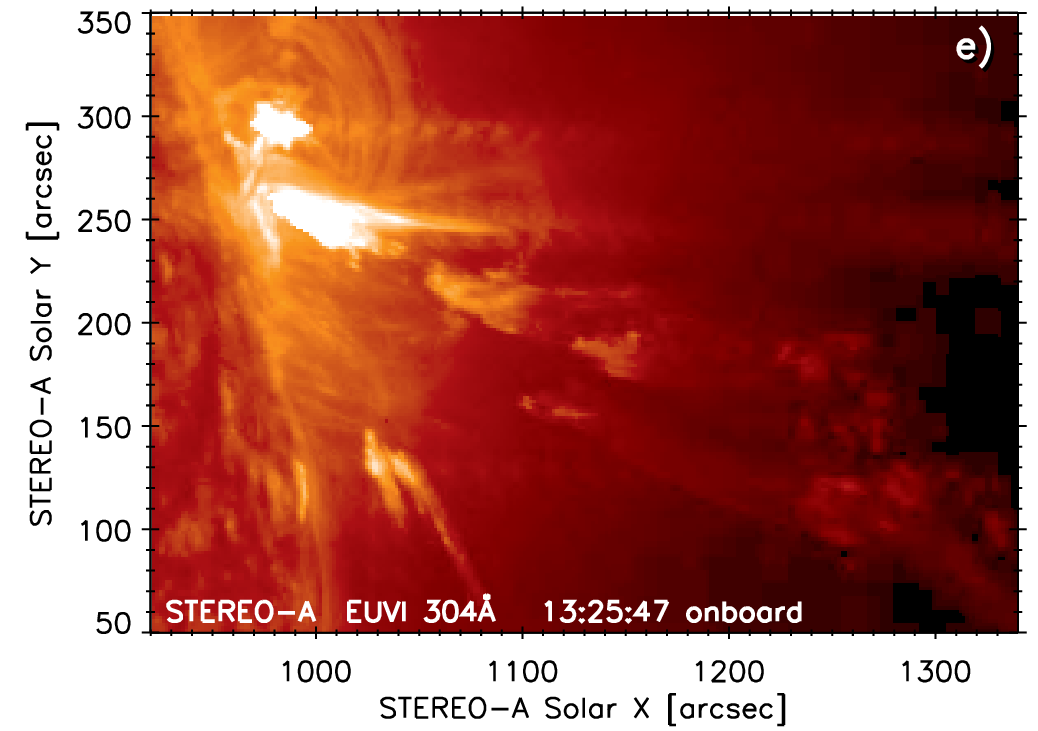}
      \includegraphics[width=6.76cm,viewport = 68  0 496 350,clip]{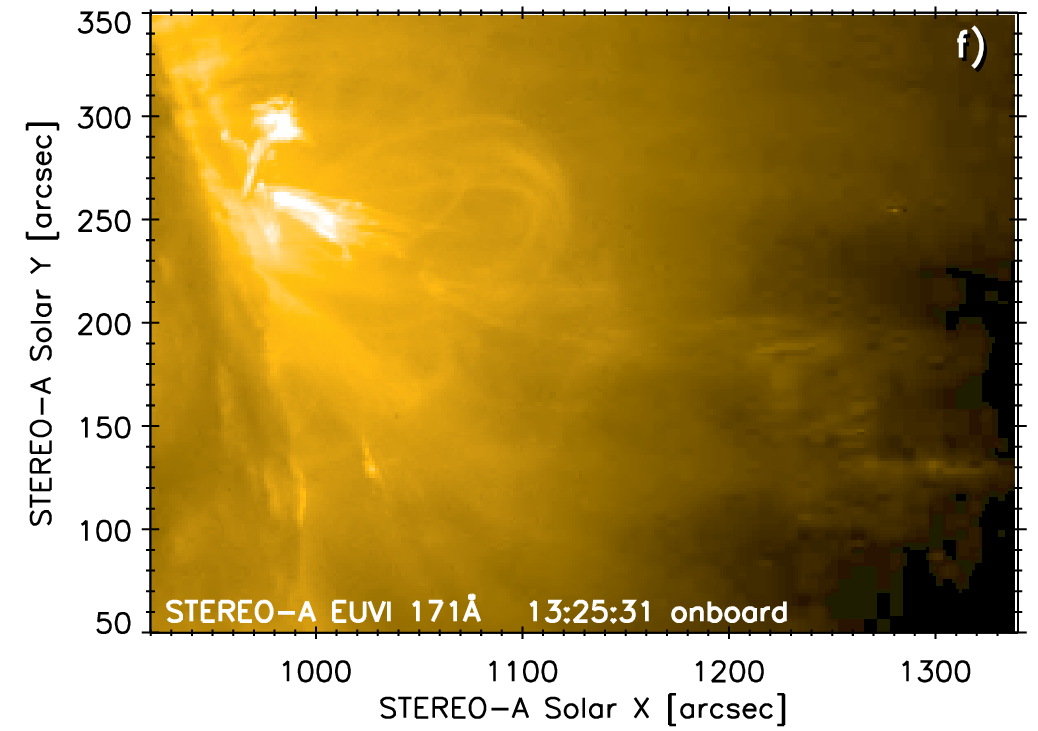}
    \end{center}
    \caption{EUV and X-ray observations of eruption around 13:25:30\,UT (STIX time interval IV). (a) flare site at 13:20:45 EUI/FSI onboard time (13:26:06\,UT). (b) and (c) SDO/AIA observations in the 171 and 304\,AA~channels, respectively. The quasi-circular feature is denoted by white arrows in (c). (d) MGN composite of the 131\,\AA~(red), 193\,\AA~(green), and 171\,\AA~(blue), animation spanning 12:30--13:55~UT is available online. Red arrow points to a hot loop observed in 131\,\AA\, along the filament leg. Contours indicate 50 and 90~\% levels of STIX sources over-plotted on EUI/FSI (a) and re-projected to AIA (c, d). The 8--11~keV sources is displayed for three assumed altitudes 0, 30, and 50 Mm in (c) but only for the 30 Mm altitude in (d). (e) and (f) complementary observations from STEREO-A/EUVI 304\,\AA~and 171\,\AA, the quasi-circular feature is detectable in both passbands, animation spanning 13:00--13:40~UT is available online.
    }
    \label{Fig:euv_1325UT}
\end{figure*}
\subsubsection{Quasi-circular feature at 13:25 UT}
\label{Sect:QCF}
The hot emission that appeared near the vicinity of the filament leg at 13:21\,UT persists at this location for at least several minutes. 
Till about 13:26 UT the northern filament leg interacts with the overlying arcade producing a system of hot flare loops 
with footpoints in a similar region as the filament leg, see AIA 131\,\AA\, and MGN images in Fig.~\ref{Fig:Hot_loop} and accompanying video there. 
At 13:25\,UT, one hot loop is located close to 
the filament leg (red arrow in Fig.~\ref{Fig:euv_1325UT}). A similar structure is present in the much cooler AIA channels of 171\,\AA~(panel b) and 304\,\AA~(panel c), but in both it is thinner and fainter. Contrary to that, in 131\,\AA, the hot loop is much broader and brighter, leading to a distinctly red loop constituting the filament leg (Fig.~\ref{Fig:euv_1325UT}d). 

The rising filament and BHF show untwisting motions and an extending helical structure starts to form around 13:23 UT at $\sim$ [900\arcsec, 250\arcsec] visible in SDO/AIA and STEREO-A/EUVI 304 and 171\,\AA\, channels, see videos accompanying Fig.~\ref{Fig:aia_overview} and Fig.~\ref{Fig:euv_1325UT}. 
For several moments  these motions lead to creation of a quasi-circular feature (QCF, Fig.~\ref{Fig:euv_1325UT}c). It consists of a series of blobs visible primarily in 304\,\AA. Only traces of it are visible in both 171\,\AA~and 131\,\AA~(Fig.~\ref{Fig:euv_1325UT}b and f) as its contrast there is low due to the background corona. The blobs occur in a quasi-circular arrangement, with two blobs in the middle of the structure. This feature is visible only in a few AIA snapshots around 13:25:30\,UT, and promptly disperses with the expanding erupting filament. The quasi-circular feature is also detected by the STEREO-A/EUVI 304\,\AA~ and 171\,\AA\, filters (Fig.~\ref{Fig:euv_1325UT}e and f STEREO animation there), with similar morphology.

At a  similar time, during 13:25:05--13:25:38\,UT, unique radio bursts 17, 18, 20, and 22 occur, see Table~\ref{Table1} and Fig.~\ref{Fig:radio_1324_1325}. The burst 17 consists of a series of positively drifting bursts followed by the burst 18 originating from almost the same frequency, $\sim$~1300~MHz, but with negative frequency drift. At 13:25:18\,UT, two most distinct bursts occur, again starting almost at the same time and frequency. The burst 20 occurs in the 1600--2000 MHz range together with the burst 22 in the 2000--2800 MHz range. They both resemble a tangle of U- and inverted U-bursts, indicating that the particles generating these bursts move in a complex magnetic structure, at first to lower densities and then to higher ones in the case of the burst 20 and, at the same time, in the opposite direction for the case of burst 22.

Additionally, these quasi-circular EUV and radio features are co-temporal with an HXR spike which occurred during the decay of the 25--50~keV emission at the STIX time interval IV, see Fig.~\ref{Fig:HXR_evolution} and Table~\ref{tab:stix_time_intervals}. STIX spectra after $\sim$13:23:45~UT, i.e. covering the STIX time intervals III, IV, and in between, are fitted well by a thermal component and a non-thermal thick-target component, which dominates energies above $\sim$15~keV, see Fig.~\ref{Fig:STIX_spectra}c and d for the time intervals III and IV, respectively. Image reconstruction at the non-thermal energies reveals sources which are interpreted as chromospheric counterparts of loop footpoints, see Fig.~\ref{Fig:stix_1318} (middle, right) located close to the hot loop and the  filament leg at this time (Figure \ref{Fig:euv_1325UT}d).

Interestingly, during the HXR spike, i.e. at the STIX time interval IV, a double power-law electron beam with a rather low value of the break energy $\sim$~25 keV was necessary to obtain a satisfactory fit with a randomly distributed residuals in energy, see Fig.~\ref{Fig:STIX_spectra}d. Different slopes, i.e. power-law indices, below and above the break energy suggest presence of two different electron distributions producing the HXR emission. This is further indicated by two different HXR sources reconstructed in two energy ranges 25--28~keV and 32--45~keV. Their position is different within 50\% contours, yet the uncertainties are much larger,  see Fig.~\ref{Fig:stix_1318} (right). Thus, we conclude there are both spectral and imaging indications of a change in electron distribution in the time interval IV, covering an HXR spike during the decay of the 25--50~keV emission. 

Reconstructed STIX sources dominated either by thermal (below 11 keV) or non-thermal X-ray emission (above 16 keV) are overlaid on EUI/FSI and reprojected to SDO/AIA images, see Fig.~\ref{Fig:euv_1325UT}. Although the reprojection into the AIA vantage point has to assume a certain height of the STIX source above the solar surface at the local normal vector, the possible AIA counterparts include filament footpoints and the system of hot loops in their vicinity but not the QCF, see reprojections of the 8--11~keV STIX thermal source assuming three heights 0,~30, and 50~Mm in Fig.~\ref{Fig:euv_1325UT}c. Yet, there could be a faint X-ray emission related to accelerated particles producing radio bursts 20 and 22, which is, however, not reconstructed as an image source due to presence of much brighter sources and limited STIX signal-to-noise ratio.
\section{Discussion}
\label{Sect:Interpretation}
We detected presence of hot loops in EUV, SXR, and STIX (deka-)keV X-ray emission in the vicinity and above the filament before its eruption. STIX spectral fitting and imaging results supported interpretation of X-ray emission as of thermal, super-hot, or coronal thick-target origin. Thus, we related those STIX sources to the hot structures along their reprojected heights above the surface, see Fig.~\ref{Fig:Hot_loop}a.

As the filament started to rise, it interacted with the overlying hot loops. The interaction is followed by a strong filament brightening, untwisting motions, and formation of EUV bright helical structure (Sects.~\ref{Sect:BHF},~\ref{Sect:QCF}), suggesting that the magnetic configuration of the erupting filament is that of a flux rope.

\begin{figure}
    \begin{center}
\resizebox{\hsize}{!}{\includegraphics{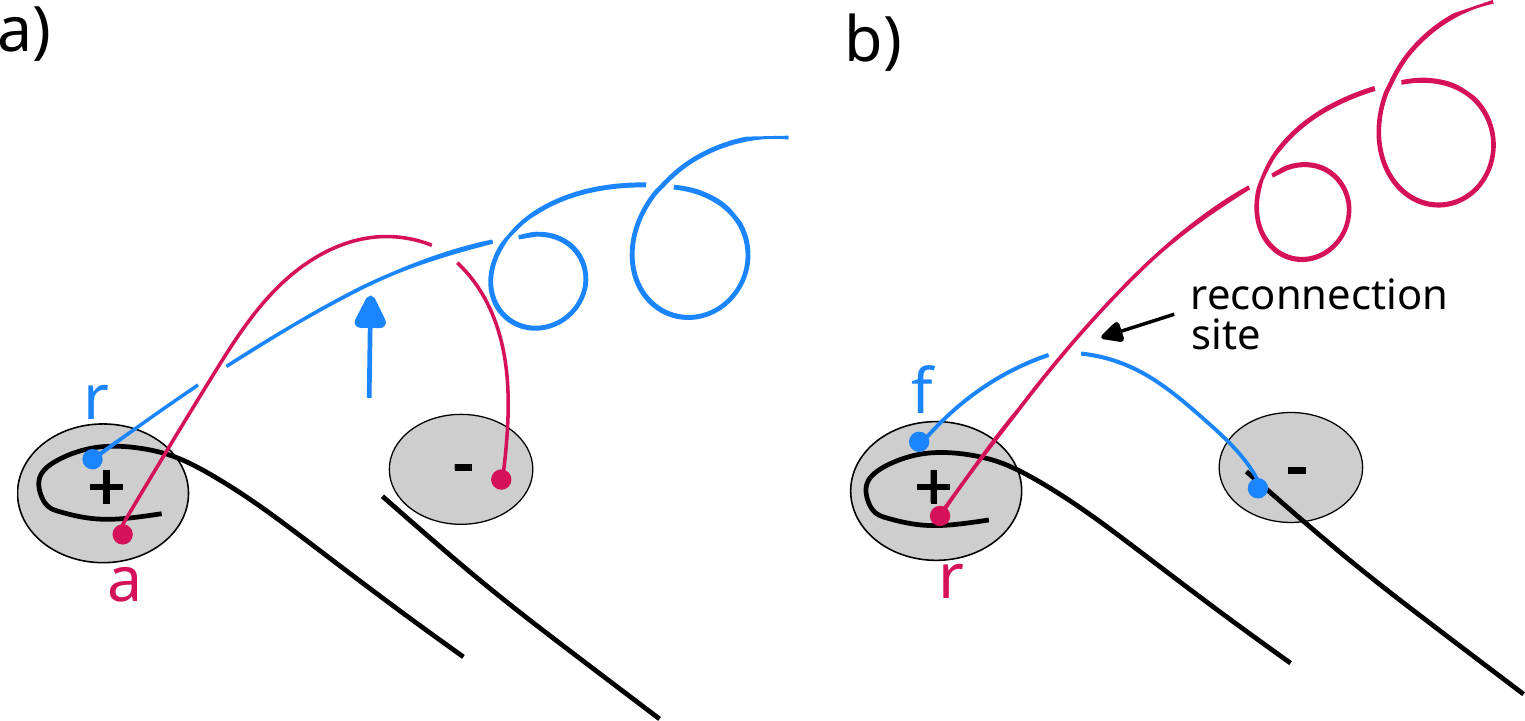}}

              \resizebox{\hsize}{!}{\includegraphics{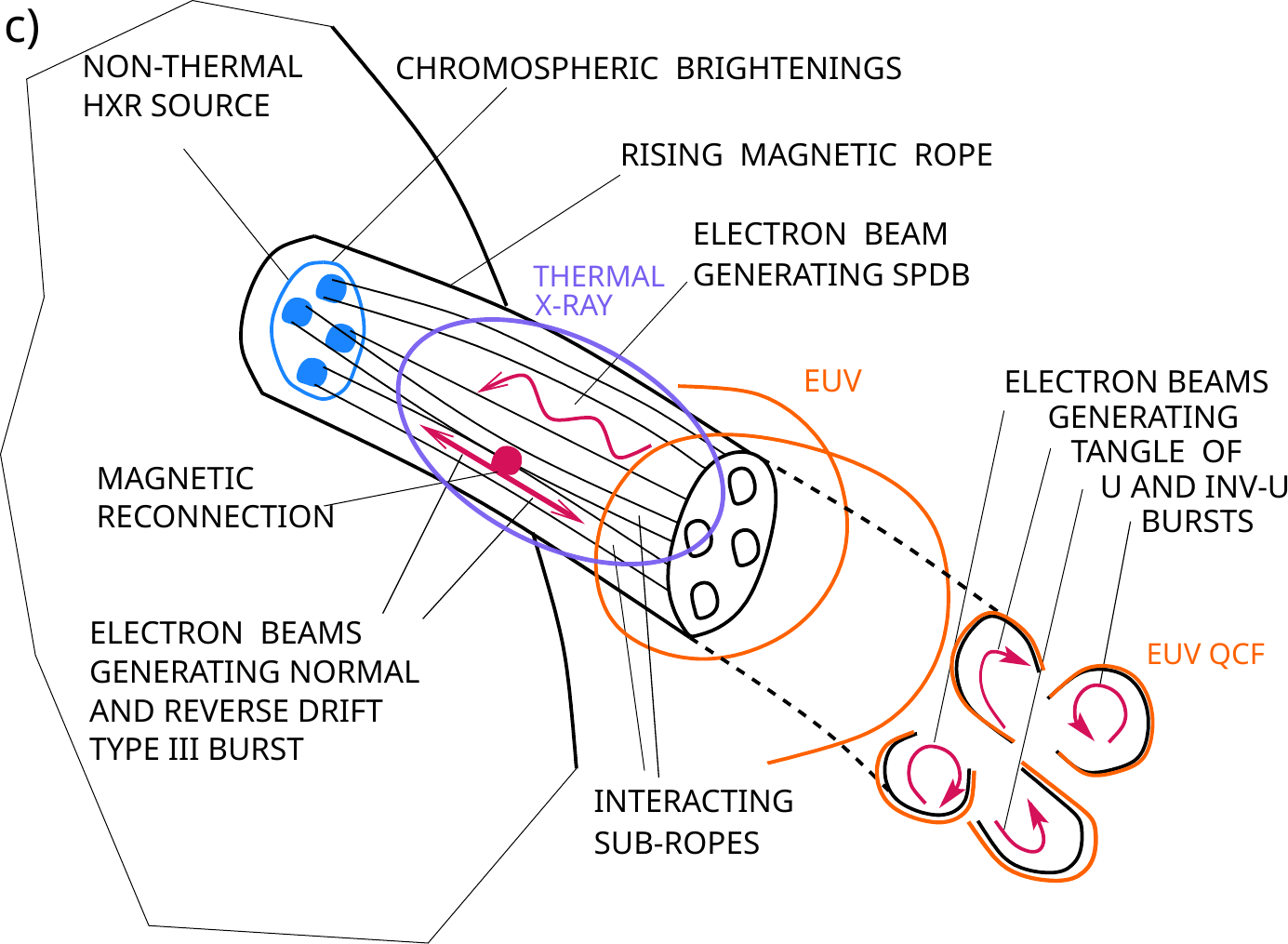}}
    \end{center}
    \caption{Schemas. Reconnection in ar--rf  geometry involving (a) arcade~`a' and flux rope~`r', producing (b) new flux rope field line~`r' and flare loop~`f'. 
    Black arrow points to an assumed reconnection site between the `r' and `f' field lines.
    (c) Processes generating radio (SPDBs, reverse and normal drift type III bursts, and tangle of U- and inverted U- bursts, Table~\ref{Table1}); indication of related EUV, thermal X-ray, and HXR emissions. Schemas are not scaled in space, regions of electron accelerations are only illustrative.
    }
    \label{Fig:schema}
\end{figure}

We interpret the interaction and the following observational evidence as signatures of reconnection in the arcade-to-rope (ar--rf) geometry, see Fig.~5 in \citet{Aulanier19} and a schema in Fig.~\ref{Fig:schema}a, b. There, an arcade loop `a' reconnects with a rising flux rope `r' (Fig.~\ref{Fig:schema}a). A new flux rope field line `r' is produced  at the footpoint of the former arcade `a' and a new flare loop `f' appears below the rope at the footpoint of previous flux rope leg `r' (Fig.~\ref{Fig:schema}b).

In this event we traced two hot arcade loops A and B overlying the filament (Fig.~\ref{Fig:Hot_loop}) which interacted/reconnected with that filament (Fig.~\ref{Fig:Hot_loop_304}a). The arcade loops A and B changed their shape and, consequently, the filament leg slipped in front of the newly formed hot loops (Fig.~\ref{Fig:Hot_loop}l) and nearby dark structures (Fig.~\ref{Fig:Hot_loop_304}c--f). Simultaneously, the filament leg brightened and the BHF occurred where the rising filament met hot loops A and B, see Fig.~\ref{Fig:Hot_loop}, panels f--h or k--m. \citet{Joshi2025} also reported  strong EUV filament brightenings related to the ar--rf type of reconnection. However, this is the first report of a hot arcade loop involved in ar--rf reconnection. The corresponding STIX (deka-)keV X-ray sources could be co-spatial either with the top part of the hot loop arcade and BHF, or the filament leg and the hot loops near it, see  different heights 0 and 50~Mm in Fig.~\ref{Fig:Hot_loop}c, d. 

Radio slowly positively drifted bursts (SPDBs) are another signature of the reconnection taking place \citep{2020ApJ...905..111Z, Zemanova2024}. We interpret bursts 1 and 2, see Fig.~\ref{Fig:radio_spdb} and Sects.~\ref{Sect:Radio_obs} and \ref{Sect:BHF}, in the same way as the SPDB in \cite{2020ApJ...905..111Z}. These bursts were observed during the appearance of the BHF. Since the acceleration process in the magnetic reconnection is typically associated with plasma heating, the BHF can be considered a site where electrons are accelerated. Based on the location of the BHF, we propose that the reconnection process occurs within the rising and unstable magnetic rope. Here, electron beams are accelerated and propagate downward along the helical structure inside the rope into the solar atmosphere, generating bursts 1 and 2 (see the schematic in Fig.~\ref{Fig:schema}). Due to the helical trajectory of the electron beams in the rope’s magnetic field, and the non-zero angle between the rope axis and the vertically stratified solar atmosphere, the frequency drift of these bursts is significantly lower compared to that of bursts generated by beams travelling along a vertical straight path. Figure~\ref{Fig:Hot_loop}h shows that the length of one helical turn along the part of the magnetic rope is much shorter than the length of the magnetic rope itself (Sect.~\ref{Sect:Evolution_EUV}), thus making the path along which the particles propagate longer than the magnetic rope itself.

X-ray spectra and sources observed during occurrence of radio bursts with a reversed frequency drift in the 0.8--4.5 GHz range were studied by \citet{2007SoPh..240..121F}. There, four bursts showed a reversed drift less than 500 MHz~s$^{-1}$. Three of them showed purely thermal X-ray spectrum and one a combination of a thermal and a non-thermal component. The temperature of thermal components was high ($\ge 25$ MK). Moreover, in two cases the thermal X-ray spectrum had two components. In the event here, the HXR emission during the radio bursts 1 and 2 was interpreted as due to thermal components, see the double thermal fit in Fig.~\ref{Fig:STIX_spectra}b and the discussion in Sect.~\ref{Sect:BHF}.
In this early flare phase,  the amount of non-thermal electrons (beam electrons) was probably too small to produce a significant non-thermal HXR component, but enough to generate SPDBs. Namely, while SPDBs are generated by the coherent emission mechanism based on the derivative of the electron  distribution function in the velocity space, HXR emission is generated by the bremsstrahlung mechanism and depends on the amount of accelerated electrons.

As the filament kept rising, the type II burst below 100~MHz related to a shock in higher corona appeared at $\sim$~13:24~UT (Fig.~\ref{Fig:radio_overview}). At higher frequencies, type III bursts generated by upwards and downwards propagating electron beams occurred (Table~\ref{Table1}). 
Co-temporal HXR emission of non-thermal origin dominated energies above 15~keV and it is interpreted as due to thick-target emission of power-law electron beams (Fig.~\ref{Fig:STIX_spectra}c). Image reconstruction resulted in one HXR source only and we assume it corresponds spatially to the strongest AIA 1600~\AA\, emission located near the footpoint of the filament (Fig.~\ref{Fig:stix_1318} middle and Sec.~\ref{Sect:X_obs}).

Then, at $\sim$~13:25~UT, rare radio bursts 17 and 18 started at nearly the same frequency ($\sim$~1300 MHz, Fig.~\ref{Fig:radio_1324_1325}). While bursts 17 drifted to higher frequencies, in the case of bursts 18 it was opposite. The frequency drifts of these bursts are relatively low, on the order of a few hundred MHz or less, indicating that the electron beams were propagating along paths significantly inclined from the vertical. The opposite frequency drifts of the bursts, along with their similar starting frequencies, suggest that the region where these bursts originate corresponds to the magnetic reconnection site itself. We interpret these bursts as being generated by electron beams travelling through the complex structure of the magnetic rope. The magnetic reconnection responsible for accelerating these beams could occur within the rope itself (note low frequency drifts of the bursts 17 and 18, similarly as in SPDBs 1 and 2, see also Fig.~\ref{Fig:schema}), or in the interaction region between the rope and an external magnetic structure as supported by the presence of a hot loop near the filament and a non-thermal X-ray source located near the filament leg.
Although not shown here, at this time the HXR source was located at the similar position as in the STIX time interval III (before) and IV (after), see Sect.~\ref{Sect:QCF} and Figs.~\ref{Fig:stix_1318} and \ref{Fig:euv_1325UT}, respectively. Recently, \citet{Huang2025} interpreted an observation of a two-sided loop jet in H$\alpha$ and EUV wavebands as driven by internal reconnection between filament threads.

The fact that both bursts 17 and 18 started at about of the same frequency allows us to estimate the plasma density in the reconnection (acceleration) region using the observed, $\sim$ 1300 MHz, starting frequency. The resulting electron density is 2.1 $\times$ 10$^{10}$ cm$^{-3}$ (emission on the fundamental frequency) or 5.2 $\times$ 10$^9$ cm$^{-3}$ (emission on the harmonic frequency).
For comparison, in \cite{2016SoPh..291.2407T}, where another example of a pair of normal and reverse type III bursts was analyzed, the frequency corresponding to the acceleration region was estimated to be around 1400 MHz, indicating plasma densities in the acceleration region similar to those in the present case.

Several seconds after the bursts 17 and 18, the unique radio bursts 20 and 22 occurred, consisting a tangle of normal U- and inverted U-bursts on the radio spectrum (Table~\ref{Table1} and Fig.~\ref{Fig:radio_1324_1325}). They were detected  simultaneously with an unusual QCF, seen in the SDO/AIA and STEREO/EUVI 304\,and 171 \AA\, (Fig.~\ref{Fig:euv_1325UT}). This structure was probably formed by interacting sub-ropes in highly twisted magnetic rope which exhibited untwisting motions (Sec.~\ref{Sect:QCF}, video accompanying Fig.~\ref{Fig:euv_1325UT}). EUV images indicate that the plasma density inside this structure was higher than that in the surrounding corona. We propose that in this structure the electron beams were accelerated and trapped. Here, the beams moved from higher to lower densities and then again to higher plasma densities, and at the same time other beams moved from lower to higher densities and then again to lower plasma densities, or even circulated there, see the schema in Fig.~\ref{Fig:schema}. These beams generated electrostatic plasma waves that produced, after their transformation into the electromagnetic ones, these bursts. 

In addition, these EUV and radio features were co-temporal with an HXR spike. Fitting of the corresponding STIX spectrum and the image reconstruction indicate there were two different beam electron distributions below and above $\sim$~30~keV (Sec.~\ref{Sect:QCF}). First, there is a spectral break near that energy in the thick-target electron spectrum (Fig.~\ref{Fig:STIX_spectra}d). Second, positions of HXR sources in the 25--28~keV and 32--45~keV are not the same within 50\% contour of their flux maximum; the centroid of the 32--45~keV source is shifted towards the direction of movement of the filament leg (Fig.~\ref{Fig:euv_1325UT}d). Although the uncertainties of the centroid positions are significant (Fig.~\ref{Fig:stix_1318} right), comparison of both spectra (Fig.~\ref{Fig:euv_1325UT}c, d) and images (Fig.~\ref{Fig:stix_1318} centre, right), respectively, suggests there is a change in the beam electron distribution probably related to the QCF. 

In summary, all the presented signatures detected at the very start of the filament rising in radio, EUV, SXR, and HXR point to the interaction of the magnetic rope with the hot overlying arcade and electron acceleration within that erupting rope. We propose their relation to magnetic rope reconnection processes.
\section{Conclusions}
\label{Sect:Concl}

We have collected indirect, yet multi-wavelength and multi-viewpoint evidence of the magnetic reconnection in the arcade-to-rope geometry (ar--rf). Before the filament eruption, hot loops were located near one filament leg. As the filament was rising, the hot loops interacted with it. The hot loops and filament exchanged connectivity (reconnected) and the filament leg slipped in front of newly formed hot flare loops. At a similar time a significant EUV brightening appeared within the erupting filament where the filament met the hot loops and radio bursts of the SPDB type occurred. Although we have no positional information of the analyzed radio bursts and the altitude of X-ray sources is uncertain due to projection effects, owing to time coincidences between radio bursts, detected helical EUV features, and proximity of X-ray sources to the leg of the erupting filament, we propose they are all related to the reconnection of a magnetic rope associated with the filament.

The helical magnetic field structure, visualised as BHF in EUV emission, prolonged the electron beam trajectory in its downward direction in the gravitationally stratified solar atmosphere and thus resulted in the slow frequency drift of the generated radio bursts. The helical structure also indicated presence of the electric current in this part of the magnetic rope. The acceleration process associated with magnetic reconnection is typically accompanied by plasma heating, which is evident from multi-thermal nature of BHF and extended X-ray sources of a spectral shape consistent with thermal origin. Therefore, the magnetic reconnection appears to occur within the magnetic rope, based on the location of the BHF.

Rising of the filament was accompanied by strong untwisting of the helical structure detected in EUV. Occurrence of normal and reverse drift type III radio bursts, which originated at a nearly same frequency, also point to a reconnection event. This radio emission allowed us to estimate electron density at the reconnection site, whereas the HXR source confirmed presence of accelerated particles associated with the erupting flux rope. Furthermore, the untwisting motions led to creation of a quasi-circular EUV feature, which was co-temporal with a unique tangle of radio U- and inverse U-bursts. We propose that accelerated electron beams moved or even circulated in this complex structure and generated the tangle of these bursts. Co-temporal HXR emission, which is interpreted in a thick-target collisional model as due to beam electrons, was concentrated near one side of hot loops and filament leg only. This further suggests that the reconnection and particle acceleration occurred near the filament leg and that the accelerated particles propagated predominantly towards the rooting of the filament in the chromosphere and not along both parts of the hot flare loops.

Therefore, we argue that at the start of this flare the main magnetic reconnection processes were not located below the rising magnetic rope. Rather, we propose that reconnection processes can be understood in terms of arcade-to-rope geometry and internal reconnections inside the rising and unstable magnetic rope. 

\begin{acknowledgements}
J.D., M.K., J.K., and A.Z. acknowledge 
the institutional support RVO:67985815 from the Czech Academy of Sciences.
A.Z. acknowledges support from the Grant GA CR No.\,22-07155S, 
J.D. acknowledges support from the Grant GA CR No. 25-18282S. M.K. and J.K. also acknowledge support from the Grant GA CR 22-34841S. J.R. acknowledges support from the Science Grant Agency project VEGA 2/0043/24 (Slovakia).
We thank D.~Berghmans for comments on EUV data and M.~Mierla for providing us with EUI data.
We also thank the Greenland-Callisto spectrum and the RSDB service at LESIA / USN (Observatoire de Paris) for making the NRH/ORFEES/NDA data available. SDO data were obtained courtesy of NASA/SDO and the AIA and HMI science teams. Solar Orbiter is a space mission of international collaboration between ESA and NASA, operated by ESA. The STIX instrument is an international collaboration between Switzerland, Poland, France, Czech Republic, Germany, Austria, Ireland, and Italy. The EUI instrument was built by CSL, IAS, MPS, MSSL/UCL, PMOD/WRC, ROB, LCF/IO with funding from the Belgian Federal Science Policy Office (BELSPO/PRODEX PEA C4000134088); the Centre National d’Etudes Spatiales (CNES); the UK Space Agency (UKSA); the Bundesministerium für Wirtschaft und Energie (BMWi) through the Deutsches Zentrum für Luft- und Raumfahrt (DLR); and the Swiss Space Office (SSO). We also acknowledge the use of the Fermi Solar Flare Observations facility funded by the Fermi GI program (\href{https://hesperia.gsfc.nasa.gov/fermi\_solar/}{https://hesperia.gsfc.nasa.gov/fermi\_solar/}).
Hinode is a Japanese mission developed and launched by ISAS/JAXA, with NAOJ as domestic partner and NASA and STFC (UK) as international partners. It is operated by these agencies in co-operation with ESA and the NSC (Norway).
\end{acknowledgements}
\bibliography{2022-04-02_Eruption_AA}
\bibliographystyle{aa}
\begin{appendix}
\section{Spacecraft positions}
\begin{figure}
    \centering
	\includegraphics[width=8.80cm,viewport = 0 0 680 520,clip]{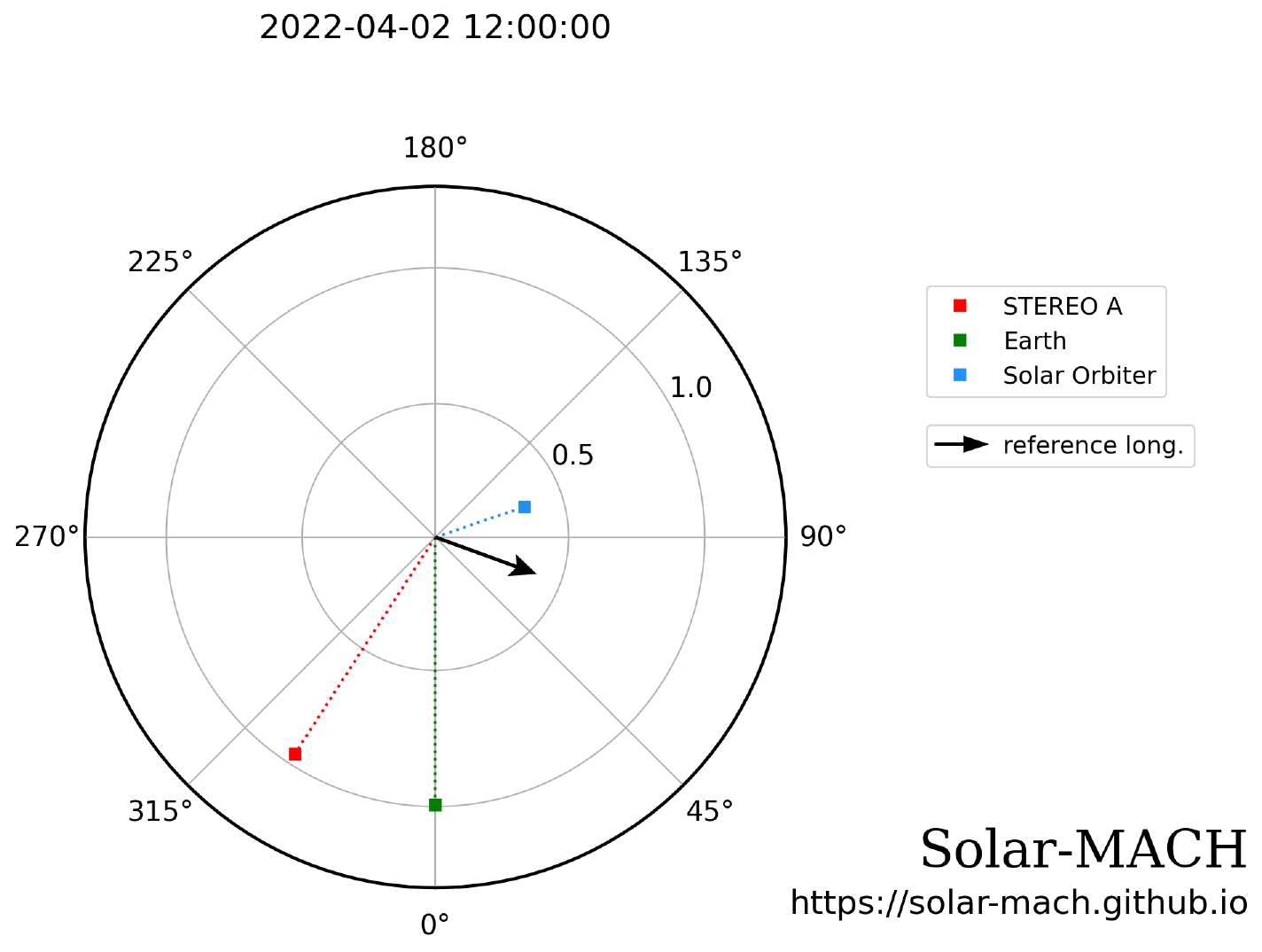}   
    \caption{Positions of spacecrafts observing flare in Stonyhurst (HEEQ) longitudes. Black arrow indicates the flare location.} 
    \label{Fig:Spacecrafts}
\end{figure}

The flare was observed from several vantage points, those data provided us with multi-directional information of the same event, see position of instruments and time differences relative to Earth in Fig.~\ref{Fig:Spacecrafts} and Table~\ref{Table2}.
\begin{table}[!ht]
  \caption{Positions of spacecrafts observing the Sun in Stonyhurst (HEEQ) coordinates and their time difference $\Delta t$ relative to Earth. Values are given for 2 April 2022, 12:00\,UT.} \label{Table2}
\begin{center}
\begin{tabular}{cccc}
  \hline\hline
                         & Earth            & STEREO-A             & Solar Orbiter     \\
  \hline
HEEQ lon          &  0.0$^\circ$     & $-$33$^\circ$    &  +109$^\circ$  \\
HEEQ lat           & $-$6.5$^\circ$   & $-$7.3$^\circ$     &    +2.7$^\circ$  \\
$\Delta t$ to Earth time    &  $-$             & 0\,min 16.51\,s        & 5\,min 21.22\,s   \\
\hline
\end{tabular}
\end{center}
\end{table}
\section{Time-distance plot}
\label{App:time_plot}
\begin{figure}[!ht]
    \centering
    \includegraphics[width=1.0\linewidth]{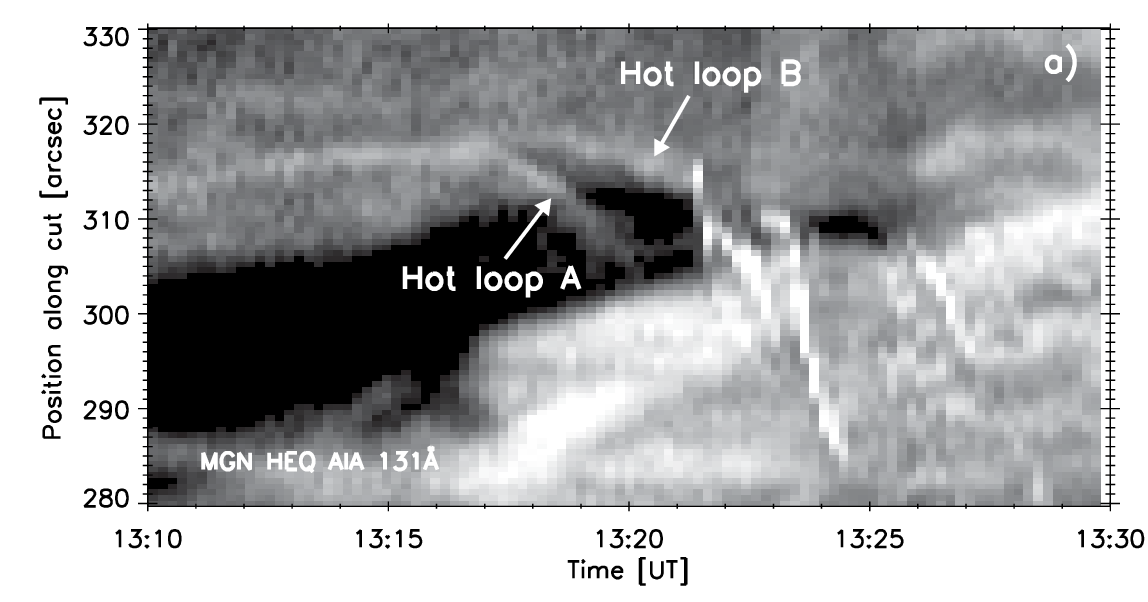}
    \includegraphics[width=1.0\linewidth]{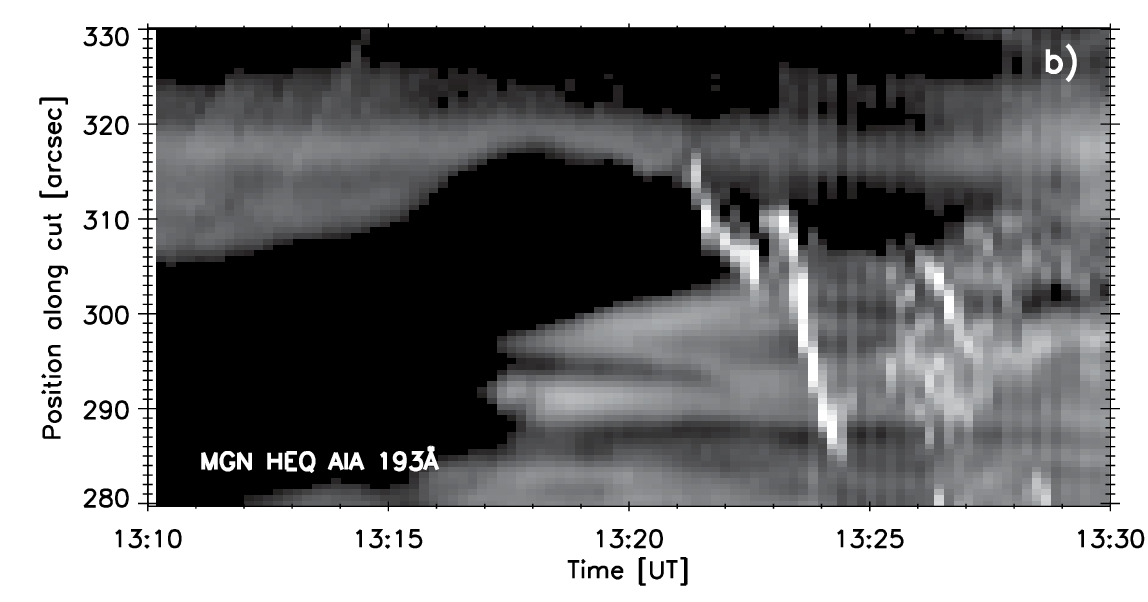}
    \includegraphics[width=1.0\linewidth]{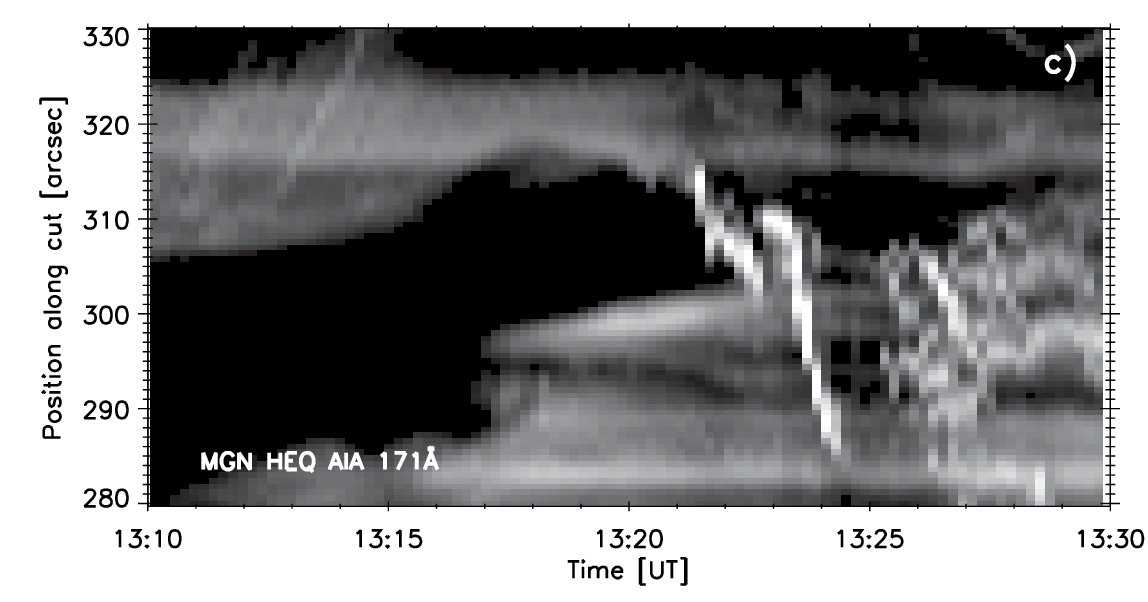}
    \caption{Separate layers of the MGN time-distance plot displayed in Fig.~\ref{Fig:Hot_loop_304}a. Panel a) shows clearly hot loops A and B in AIA 131~\AA. Panels b) and c) do not show hot loop A at all in AIA 193 and 171~\AA, respectively, instead background coronal emission is visible near the position of the hot loop B.}
    \label{fig_stck_layers}
\end{figure}
Although 3-colour images using three SDO/AIA filters help distinguish hot structures emitting predominantly in 131~\AA, see Sect.~\ref{Sect:Hot_plasma}, mixture of several structures along the line of sight cannot by always avoided. This is the case of the hot loop B. In MGN 3-colour image, the loop B can be traced as a thin red loop, see Fig.~\ref{Fig:Hot_loop}k, whereas in the time-distance plot along the cut, the loop is of rainbow colour, see Fig.~\ref{Fig:Hot_loop_304}a. Time-distance plots of separate MGN layers, see Fig.~\ref{fig_stck_layers}, reveal that the loop B is clearly visible in emission in 131~\AA\, similarly as the hot loop A. The emission visible in 193 and 171~\AA\, belongs to an extended coronal structure located behind the hot loop B, compare Fig.~\ref{Fig:Hot_loop} and panels b and c in Fig.~\ref{fig_stck_layers}.
\section{X-ray sources at low energies}
\label{App1}
\begin{figure*}
\includegraphics[width=\textwidth]
{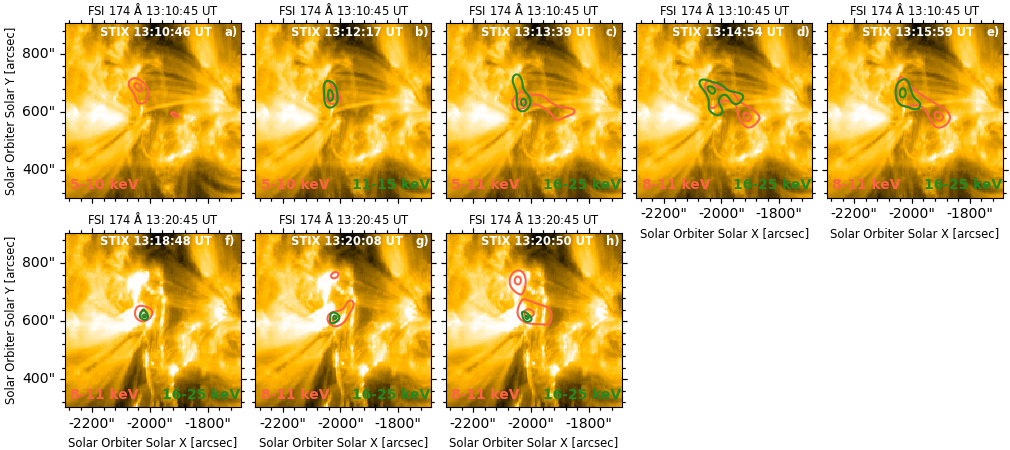}
\caption{Early evolution of STIX X-ray sources overlaid on two EUI/FSI 174\AA\, images. Contours indicate 50 and 90 \% levels of the peak flux in the STIX images, all times on images are onboard times. Top row: X-ray sources of possible thermal origin up to 25 keV which are close to the FSI image taken at 13:10 onboard time (13:16 UT). Bottom row:  subsequent times, closer to the FSI image taken at 13:20 onboard time (13:25 UT), when the 16--25 keV source is of non-thermal origin.}
\label{fig_eui_stix_evol}
\end{figure*}

STIX X-ray emission above 5~keV related to the studied loop system started to rise at $\sim$~13:16~UT. The STIX spectrum could be fitted reasonably well by a thermal and a very steep thick-target component. Imaging of the thermal component in the 5--10~keV range revealed an elongated source near eastern side of the loop system seen in EUI/FSI 174~\AA\, image, see Fig.~\ref{fig_eui_stix_evol}a. If a height of 50 Mm above the local surface is assumed, that source can be co--aligned with a top part of a hot loop system visible on Be thin XRT image. Later on, at $\sim$ 13:17 UT (Fig.~\ref{fig_eui_stix_evol}b), it is possible to reconstruct a 11--15~keV source located at the same position as the 5-10 keV thermal source. Assuming again 50 Mm height both sources can be co-aligned with hot loop system detected by XRT. The evolution of such sources, i.e. of possible thermal origin, and their relation to XRT emission is displayed in Fig.~\ref{Fig:Hot_loop} (top row, panels a--d). Fig.~\ref{fig_eui_stix_evol} shows the evolution from vantage point of Solar Orbiter and without assumption on source height. 

The thermal 5-11~keV source at the STIX time interval III is significantly smaller than the thermal source at the STIX time interval II (compare panels e) and f) of Fig.~\ref{fig_eui_stix_evol}). 
Between the STIX time intervals III and IV the thermal sources grow larger with time, although not reaching the extent of the thermal source in the time interval II, see panels f and g of Fig.~\ref{fig_eui_stix_evol}). Temperature of the thermal component stays similar, $\sim$1.5~keV over time intervals III and IV, the emission measure rises.
The panel h) of Fig.~\ref{fig_eui_stix_evol} shows STIX sources corresponding to the time interval of EUI/FSI 174~\AA\, image at 13:20 onboard time.
\end{appendix}
\end{document}